\documentclass[twocolumn]{aastex63}
\usepackage{amsmath}
\usepackage{amsfonts}
\usepackage{amssymb}

\usepackage{color}

\usepackage{multirow}

\usepackage{comment}
\usepackage{relsize}

\usepackage{lineno}

\newcommand{\Mmax}{\ensuremath{M_\mathrm{max}}}
\newcommand{\Mtov}{\ensuremath{M_\mathrm{TOV}}}
\newcommand{\Mform}{\ensuremath{M_\mathrm{form}}}
\newcommand{\Msolar}{\ensuremath{M_\odot}}

\newcommand{\EOS}{\ensuremath{\mathrm{EOS}}}

\newcommand{\data}{\ensuremath{\mathcal{D}}}
\newcommand{\detected}{\ensuremath{\mathrm{det}}}

\newcommand{\Hyp}{\ensuremath{\mathcal{H}}}

\newcommand{\result}[1]{\ensuremath{#1}}

\begin{document}

\title{
Discriminating between Neutron Stars and Black Holes with Imperfect Knowledge of the Maximum Neutron Star Mass
}

\author{Reed Essick}
\email{reed.essick@gmail.com}
\affiliation{Kavli Institute for Cosmological Physics, The University of Chicago, Chicago, IL 60637, USA}

\author{Philippe Landry}
\email{plandry@fullerton.edu}
\affiliation{Gravitational-Wave Physics \& Astronomy Center, California State University, Fullerton, 800 N State College Blvd, Fullerton, CA 92831}

\begin{abstract}
Although gravitational-wave signals from exceptional low-mass compact binary coalescences, like GW170817, may carry matter signatures that differentiate the source from a binary black hole system, only one out of every eight events detected by the current Advanced LIGO and Virgo observatories are likely to have signal-to-noise ratios large enough to measure matter effects, even if they are present.
Nonetheless, the systems' component masses will generally be constrained precisely.
Constructing an explicit mixture model for the total rate density of merging compact objects, we develop a hierarchical Bayesian analysis to classify gravitational-wave sources according to the posterior odds that their component masses are drawn from different subpopulations.
Accounting for current uncertainty in the maximum neutron star mass, and adopting a power law mass distribution with or without a mass gap and either random or mass-ratio dependent pairing, we examine two recent events from the LIGO-Virgo Collaboration's third observing run, GW190425 and GW190814.
For population models with no overlap between the neutron star and black hole mass distributions, we typically find that there is a \result{\gtrsim 70\%} chance, depending on the exact population model, that GW190425 was a binary neutron star merger rather than a neutron-star--black-hole merger.
On the other hand, we find that there is a \result{\lesssim 6\%} chance that GW190814 involved a slowly spinning neutron star, regardless of our assumed population model.
\end{abstract}


\section{Introduction}
\label{sec:introduction}

Most of the compact binary coalescences observed during the first two observing runs of the Advanced LIGO~\citep{AdvancedLIGO} and Virgo~\citep{AdvancedVirgo} detectors (O1 and O2) were neatly categorized as binary black hole (BBH) systems based on their inferred masses, which comfortably exceed the maximum neutron star (NS) mass of 3--4$\,\Msolar$ derived from basic causality arguments~\citep{RhoadesRuffini1974, VanOeveren2017}.
However, during the detectors' third observing run (O3), the LIGO-Virgo Collaboration (LVC) reported a number of low-mass systems that cannot be categorized so easily.
GW190425, a compact binary coalescence with a total mass of approximately $3.4\,M_\odot$, was deemed a likely binary NS (BNS) merger because its component masses lie between 1.12 and $2.52\,M_\odot$~\citep{GW190425}, compatible with many models of NS structure.
Although it is very likely that this system was a BNS, since no unequivocal matter effects were discernible in the gravitational wave (GW) signal and no electromagnetic (EM) counterpart was identified, there is no definitive proof that the system contained a NS~\citep{HanTang2020,KyutokuFujibayashi2020}.
Similarly, GW190814 \citep{GW190814} has a secondary mass of 2.5--2.7$\,M_\odot$, potentially consistent with either a NS or a black hole (BH).
Even the nature of GW170817~\citep{GW170817}, the seminal discovery from O2, is somewhat ambiguous~\citep{GW170817-Props, GW170817-ModelSelection, Essick2020, HindererNissanke2019, CoughlinDietrich2019, GW170817-MMA, GW170817-KN}.

Accurately classifying compact binary coalescences is important for studies of NS matter, the interpretation of electromagnetic counterparts, and inferences of subpopulation properties.
For example, mistaking BHs for NSs can bias our knowledge of the NS equation of state~\citep{YangEast2018,Chen2020}.
Kilonova models, including those used to estimate the amount of dynamical ejecta associated with GW170817~\citep{GW170817-KN}, depend partly on whether the system is a BNS or a neutron-star--black-hole binary~\citep[e.g.,][]{FernandezFoucart2017,BarbieriSalafia2019}.
Likewise, certainty that GW190425 was a BNS would alter the known properties of the distribution of NS masses in binaries because it is a strong outlier compared to known Galactic BNSs~\citep{GW190425,GuptaGerosa2020}.

Many authors have examined how well one can distinguish neutron-star--black-hole (NSBH) coalescences from BNS or BBH mergers based on their GW signals in second-~\citep{HannamBrown2013,Littenberg2015,Mandel2015,YangEast2018, JohnsonMcDanielMukherjee2018,ChenChatziioannou2020,TsokarosRuiz2020,DattaPhukon2020} and third-generation detector networks~\citep{KrishnenduSaleem2019,ChenJohnsonMcDaniel2020,Fasano2020}.
These studies rely on various matter signatures in the waveform to discern the presence of a NS in the binary.
Chief among these are tidal effects that imprint on the waveform during the inspiral.
This manifests as a phase offset relative to an equivalent BBH system caused by the integrated effects of stationary \citep[e.g.,][]{PhysRevD.77.021502, PhysRevD.79.124032}, dynamical \citep[e.g.,][]{1994MNRAS.270..611L, 1994ApJ...426..688R, PhysRevLett.116.181101, PhysRevD.94.104028}, and non-linear tides \citep[e.g.,][]{Weinberg_2016, PhysRevD.94.103012, GW170817NonlinearTides}.
The dominant part of the phase shift is due to the quasi-static equilibrium tide, parameterized by the binary tidal deformability,
\begin{equation}
    \tilde{\Lambda} = \frac{16}{13} \frac{(m_1+12m_2)m_1^4 \Lambda_1 + (m_2 + 12m_1)m_2^4 \Lambda_2}{(m_1+m_2)^5}
\end{equation}
for component masses $m_1 \geq m_2$ (by convention) with tidal deformabilities $\Lambda_1$ and $\Lambda_2$, respectively.
It is this combination of individual tidal parameters to which GW detectors are most sensitive.
However, for very asymmetric NSBHs ($\Lambda_1=0$, $m_2\ll m_1$) or massive BNSs ($\Lambda_1$, $\Lambda_2 \rightarrow 0$), $\tilde\Lambda$ can become very small, so as to be pragmatically indistinguishable from a BBH ($\tilde\Lambda=0$).
This poses a not-insignificant limitation on the use of tides to distinguish NSs from BHs~\citep{YangEast2018,JohnsonMcDanielMukherjee2018,ChenChatziioannou2020,TsokarosRuiz2020,ChenJohnsonMcDaniel2020, Fasano2020}, as the systems about which we are most uncertain are precisely those for which tidal deformations are difficult to resolve.

Several as-yet unmeasured waveform signatures could also potentially discriminate between NSs and BHs in merging compact binaries.
An abrupt truncation of the GW waveform, the unmistakable sign of tidal disruption~\citep{ShibataTaniguchi2008}, would be a clear hallmark of a NSBH.
The lack of an EM counterpart may also distinguish between a BBH and similar NSBH or BNS systems~\citep{FoucartHinderer2018, Coughlin2019, BarbieriSalafia2020}.
However, this requires a robust prediction for the amount of luminous matter left outside the merger remnant, and if the primary mass is sufficiently large or the BH spin is sufficiently small, the lighter companion could be swallowed whole.
Modifications of the GW inspiral due to a spin-induced quadrupole moment~\citep{KrishnenduArun2017,KrishnenduSaleem2019} or tidal heating~\citep{DattaPhukon2020} can also distinguish BHs from NSs, at least in principle, but it is unclear whether these effects will be measurable with second- or third-generation detectors. Future detector networks might also be able to distinguish the nature of the pre-merger components through direct observations of the post-merger remnant~\citep{TsokarosRuiz2020}.

The issues with classification schemes based on GW matter signatures are compounded by the fact that most detections made with the current advanced detectors will be near the detection threshold, and therefore matter signatures may be difficult to resolve \citep{Lackey2015}.
Indeed, 87.5\% of events detected above a threshold signal-to-noise ratio (S/N) of 10 are expected to have S/N $<$ 20, the threshold beyond which tidal effects are generally well-measured.
The case may be more optimistic for third-generation detectors \citep[e.g.,][]{Vitale2016, Haster2020}, but there will always be a population of poorly resolved events at the sensitivity threshold.
EM counterparts may help distinguish between types of binaries~\citep{YangEast2018,HindererNissanke2019,HannamBrown2013, Margalit_2019}, but these may not always be detectable, particularly for broadly localized, distant sources~\citep{Coughlin2019}.

In any case, we can reliably expect the component masses to be better constrained by GW data than the tidal deformabilities because they enter in the waveform at lower post-Newtonian order~\citep{PhysRevD.77.021502, PhysRevD.79.124032}.
Fortunately, even without any tidal information, the component masses are still informative.
Several authors have already investigated mass-based classification schemes.
Following initial work by \citet{HannamBrown2013}, \citet{Littenberg2015} quantified the typical uncertainty in mass measurements and whether posterior credible intervals will likely be small enough to confidently distinguish between NSs and BHs.
Specifically, they found that objects with true masses $\lesssim 1.4\, M_\odot$ should be confidently identified as NSs (i.e., below an assumed mass gap of 3--5$\,M_\odot$) while objects with masses $\gtrsim 6\, M_\odot$ should be confidently identified as BHs.
\citet{Mandel2015} explored the ability to distinguish between NSs and BHs with masses based on credible intervals assuming a single expected mass distribution from population synthesis calculations.
Additionally, \citet{10.1093/mnras/stw2883} explored model-independent clustering schemes to identify different types of compact object mergers based on their masses, validating their approach by recovering population synthesis predictions with simulated measurement uncertainty.
Each of these studies only considered mass distributions with large mass gaps between NSs and BHs, and they used posterior credible intervals instead of full posterior distributions.

\citet{Kapadia_2020} implemented a multi-class classification scheme based on component masses and spins inferred from a template-based GW search.
This was used to compute probabilities that individual GW events were astrophysical in origin for the LVC's GWTC-1 catalog \citep{GWTC-1}.
However, they assumed component-mass distributions that were not informed by the observed set of detections and chose fixed boundaries between classes \textit{a priori}.
Classification updates released by the LVC during O3 \citep{userguide, gracedb}, based on parameter estimation from GW data, used the full mass posteriors, although they still employed fixed boundaries between classes.

In this paper, we implement a mass-based classification for GW sources.
While the uncertainty in the compact objects' masses plays a decisive role in our approach, we directly derive posterior odds based on the full component mass posteriors.
We generalize \citet{Kapadia_2020} and \citet{userguide} by giving a complete treatment of the mass distributions, accounting in particular for uncertainty in the maximum NS mass.

We present a hierarchical Bayesian model selection scheme to determine whether individual compact objects are NSs or BHs based on their masses.
We show how to self-consistently infer the mass distribution while updating our knowledge of individual events.
For concreteness, we apply our method to two recent GW events of ambiguous nature, GW190425 and GW190814.
We restrict our population models to a few limiting cases in order to show the classification's sensitivity to the assumed mass distribution.
Our examples are not meant as a comprehensive census of proposed low-mass distributions, but rather capture the effects of a few common phenomenological features.
Additionally, we describe the types of uncertainty which limit the inference.
While Bayesian classification schemes of this kind are applicable to other problems in astrophysics, such as x-ray binaries~\citep[e.g.][]{GopalanVrtilek2015}, here we provide a treatment tailored specifically for GW sources.

Of tantamount importance to our analysis is the uncertainty in $\Mmax$, the maximum gravitational mass a NS can attain. 
For non-rotating stars, the maximum mass ($\Mtov$) is set by the equation of state (\EOS) of NS matter, which determines the internal pressure gradients that oppose gravitational collapse.
As we do not know the \EOS~perfectly, there remains considerable uncertainty in \Mtov.
In general, rotation can support more massive stars against collapse so that \Mmax~is a function of both \Mtov~and the object's dimensionless spin $\chi=cS/Gm^2$, where $S$ is the object's spin angular momentum, with $\lim_{\chi\rightarrow0} \Mmax = \Mtov$ \citep[e.g.,][]{Breu2016}.
Specifically, we approximate the effects of solid-body rotation; differential rotation can temporarily support stars of even higher masses but is not stable for viscous fluids.

Additionally, astrophysical formation channels, including any accretion or spin-up after the NS is born in a core-collapse supernova, may only produce NSs up to a mass scale $\Mform \leq \Mmax$.
Although we can still take \Mmax~as a reasonable upper bound without asserting detailed knowledge of such formation channels, we are still sensitive to the (somewhat unknown) shape of the population of merging compact objects, along with our uncertainty in \Mmax.

We briefly summarize our main conclusions in Section~\ref{sec:summary} before presenting our hierarchical Bayesian methodology in Section~\ref{sec:methods}.
Section~\ref{sec:sensitivity to population parameters} describes how we incorporate (uncertain) knowledge of the distribution of low-mass stellar remnants and the NS maximum mass in our inference.
Section~\ref{sec:examples} applies our methodology to classify GW190425 and GW190814 in greater detail.
In Section~\ref{sec:mass gap}, we explore how a definite source classification for an event can improve our knowledge of its properties, and those of the compact object population.
Finally, we conclude in Section~\ref{sec:discussion}.

\subsection{Executive summary}
\label{sec:summary}

\begin{figure}
    \begin{center}
        \includegraphics[width=\columnwidth, clip=True, trim=0.1cm 0.3cm 0.2cm 0.2cm]{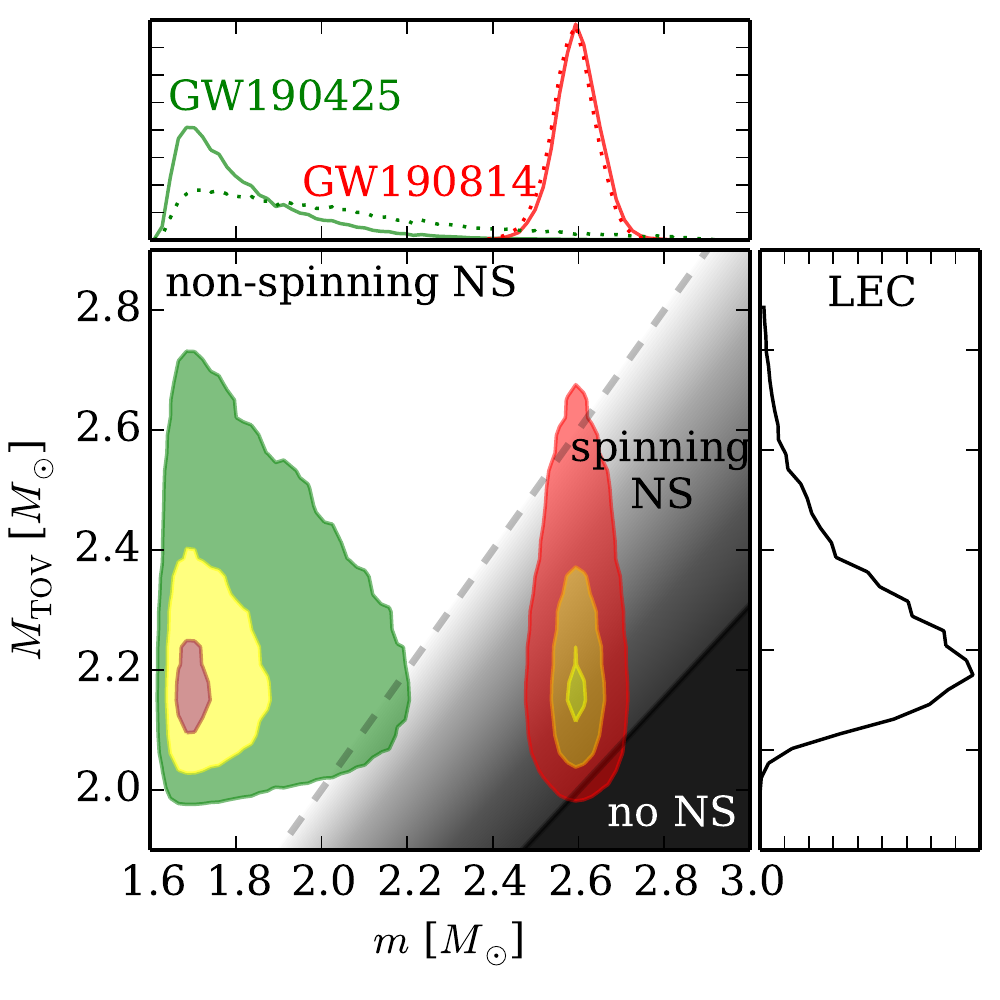}
    \end{center}
    \caption{
        Joint posterior distributions for component masses and \Mtov.
        GW190425's $m_1$ is shown in green and GW190814's $m_2$ is shown in red, assuming a power-law mass distribution with no mass gap similar to \citet{PickyPartners} ($p_\textsc{pl}(m)$ in Table~\ref{tab:models}) is relevant at low masses.
        The top panel compares the mass posteriors obtained with this assumed population (solid lines) against those for a uniform mass distribution (dotted lines).
        The \Mtov~distribution, shown in the right panel, is taken from the NS EOS inference of \citet{Landry2020} (LEC).
        In the joint distributions, the unshaded region where $\Mtov \geq m$ supports non-spinning NSs, although this region could also contain other types of stellar remnants.
        Therefore, the posterior probability within this region is an upper limit on the probability that the object is a non-spinning NS.
        Grey shading indicates the region where NSs can only exist if they are spinning, with darker shading indicating the need for larger spins.
        The darkest shaded region below the solid black line corresponds to regions where no stable NS can exist.
    }
    \label{fig:pedagogy}
\end{figure}

Before laying out our formalism in detail, we briefly summarize the results of our case studies.
We classify the binary coalescences GW190425 and GW190814 as BNSs, NSBHs or BBHs on the basis of their component masses, assuming that the NS and BH mass distributions do not overlap while accounting for uncertainty in both \Mmax~and the overall astrophysical distribution of component masses.
We take each component mass's overall prior distribution to be either uniform, a power law, or a power law with a mass gap, and assume either a random or mass-ratio dependent pairing function.
Additionally, we consider \Mmax~distributions inferred from \EOS~and population studies of NSs.
This is illustrated in Fig.~\ref{fig:pedagogy}, which shows how the $m$-$\Mtov$ parameter space is partitioned into NS-compatible and NS-incompatible regions.

The existence of $\sim 2 \, M_\odot$ NSs implies that GW190425's secondary component of $1.12$--$1.68\, M_\odot$ is a NS, while its primary component of $1.61$--$2.52 \, M_\odot$ could be either a NS or a BH, depending on \Mmax.
We quantify this by calculating $P(m_1 \leq \Mmax)$ for different uncertain estimates for \Mmax~and choices of the unknown compact object mass distribution.
The different scenarios and their resulting probabilities are given in Table~\ref{tab:GW190425}.
We find the probability that GW190425's primary is a NS to be generally $\gtrsim 70\%$, no less than $60\%$, and often $\gtrsim 90\%$.
Hence, GW190425 is likely a BNS on the basis of its component masses.

Similarly, GW190814's $\sim 23 \, M_\odot$ primary component is definitely a black hole, while the nature of its $2.50$--$2.67\,M_\odot$ secondary component is ambiguous.
Calculating $P(m_2 \leq \Mmax)$ for the same scenarios as above, we arrive at Table~\ref{tab:GW190814}.
The results are essentially insensitive to the assumed population model.
For the \Mtov~estimates from \EOS~studies, there is a $\lesssim 6\%$ chance that the secondary is a slowly-spinning NS, while for the population-based \Mmax~estimate, the probability rises to $\sim 30\%$.
Only if we assume the secondary is spinning at the breakup frequency for a NS, and that the maximum achievable NS mass is consequently boosted by rotational support, does the probability favor a NS.
Thus, we conclude that GW190814 is more likely to be a BBH than a NSBH.

\section{Distinguishing compact objects by their masses}
\label{sec:methods}

Consider the subset of GW signals for which matter signatures, like tidal deformations, are inconclusive.
We are then left with information about objects' masses as the primary way to distinguish between different types of compact binaries.

Fundamentally, we will ask whether an observed mass $m$ is most consistent with one of several known mass distributions, all of which contribute to a mixture model representing the overall rate density of compact objects such that
\begin{equation}
    \frac{d \mathcal{N}_\mathrm{tot}}{d\mu} = \sum_{\alpha} \frac{d \mathcal{N}_\alpha}{d\mu} .
\end{equation}
Greek indices label the different subpopulations (e.g., $m_1$ is a BH and $m_2$ is a BH), Latin indices label the different observed systems, and $\mu$ stands for all single-event parameters, such as component masses $m_{1,2}$ and spins $\chi_{1,2}$, with the understanding that $d\mu = dm_1 dm_2 d\chi_1 d\chi_2 \cdots$.

It is common practice to parameterize the rate density as an overall rate $\mathcal{R}$ times an overall mass distribution $p(\mu|\Theta,P)$, which itself is constructed as a mixture model.
That is,
\begin{equation}
    \frac{d \mathcal{N}_\mathrm{tot}}{d\mu} = \mathcal{R} \, p(\mu|\Theta,P) 
\end{equation}
where
\begin{gather}
    p(\mu|\Theta, P) = \sum\limits_\alpha p(\mu|\theta_\alpha) p_\alpha , \label{eq: massprior} \\
    p_\alpha = \frac{\int d\mu\, (d\mathcal{N}_\alpha/d\mu)}{\mathcal{R}} = \frac{\mathcal{R_\alpha}}{\sum\limits_\beta \mathcal{R}_\beta} .
\end{gather}
Each subpopulation is described by a mass distribution $p(\mu|\theta_\alpha)$ determined by population parameters $\theta_\alpha$ and a prior probability $p_\alpha$ equivalent to its fractional contribution to the overall rate ($\mathcal{R}_\alpha/\mathcal{R}$).
$\Theta$ represents the union of $\theta_\alpha\ \forall\ \alpha$, and $P$ represents the union of all $p_\alpha$.
We further assume all probability distributions are normalized.

In this paper, we compare posterior probabilities that a given GW event hails from different subpopulations.
We consider binaries containing only NSs and BHs, although additional populations of hypothesized exotic compact objects~\citep[see e.g.][]{CardosoPani2019} could easily be accommodated. 
Our analysis includes correlations between the masses and spins of binary components in GW mergers, as there is both theoretical~\citep[e.g.,][]{10.1117/12.458968, PhysRevD.100.043027, PhysRevD.98.123005} and empirical~\citep{GWTC1RatesAndPopulations, PickyPartners} evidence that binaries do not form through random pairings.

\subsection{Compact object classification}
\label{sec:single}

Suppose we have independent observations of $N$ systems constituting a set $\mathcal{E}$, each with separate single-event parameteters $\mu_i$ such that $\{\mu\}_\mathcal{E} = (\mu_1,\, \mu_2,\, \cdots,\, \mu_N)$.
The nature of the $k^{\rm th}$ system is unclear because of the uncertainty in both $\mu_k$ and the subpopulations.
We account for both kinds of uncertainty by explicitly modeling them within our analysis.
To wit, we construct a joint distribution over all the data $\{\data\}_\mathcal{E}  = (\data_1,\, \data_2,\, ...,\, \data_N)$, single-event parameters $\{\mu\}_\mathcal{E}$, and population parameters $(\Theta,\, P,\, \mathcal{R})$ \citep[e.g.,][]{1995ApJS...96..261L, 2004AIPC..735..195L, PhysRevD.91.023005, 10.1093/mnras/stz896, GWTC1RatesAndPopulations}:
\begin{align} \label{eq: joint}
    p( \{\mu\}_\mathcal{E}, \mathcal{R}, \Theta, P, \{\data\}_\mathcal{E}) = \; &p(\mathcal{R}, \Theta, P) \mathcal{R}^N e^{-\mathcal{R}\beta(\Theta, P)} \nonumber\\
     &\times \prod\limits_{i\in\mathcal{E}}^N p(\data_i|\mu_i) p(\mu_i|\Theta, P) ,
\end{align}
where
\begin{equation}
    \beta(\Theta, P) = \int d\mu\, P(\detected|\mu) p(\mu|\Theta, P)
\end{equation}
is the probability of making a detection given the population model, which accounts for selection effects (i.e.~observational biases) within the survey.
$P(\detected|\mu)$ is the probability of detecting a system with parameters $\mu$, and $p(\mathcal{R}, \Theta, P)$ represents our prior beliefs about all population-level parameters.
Conditioning Eq.~\eqref{eq: joint} on the observed data yields a posterior distribution, and marginalizing over the overall rate and population parameters yields population-informed posterior distributions for the observed systems' parameters $\{\mu\}_\mathcal{E}$.
Specifically,
\begin{align} \label{eq: masspost}
    p(\{\mu\}_\mathcal{E}|\{\data\}_\mathcal{E}) & = \frac{\int d\mathcal{R} d\Theta dP\, p(\{\mu\}_\mathcal{E},\mathcal{R}, \Theta, P, \{\data\}_\mathcal{E}) }{\int d\mathcal{R} d\Theta  dP ( \prod\limits_{i\in\mathcal{E}} d\mu_i ) \, p(\{\mu\}_\mathcal{E},\mathcal{R}, \Theta, P, \{\data\}_\mathcal{E})} .
\end{align}
A population-informed posterior distribution $p(\mu_k|\{\data\}_\mathcal{E})$ for $\mu_k$ follows from a marginalization over the other $N-1$ systems.

Similarly, we can compute the joint posterior for any individual system $k$ to come from a subpopulation $\alpha_k$ by extracting the part of the overall mixutre model prior for $\mu_k$ associated with $\alpha_k$:
\begin{widetext}
\begin{align}\label{eq:full posterior prob}
    p(\alpha_k|\{\data\}_\mathcal{E}) = \frac{\int d\mathcal{R} d\Theta dP\, p(\Theta, P, \mathcal{R}) \mathcal{R}^N e^{-\mathcal{R}\beta(\Theta,P)} \left( \prod\limits_{i \neq k} d\mu_i \, p(\data_i|\mu_i) p(\mu_i|\Theta, P) \right) d\mu_k\, p(\data_k|\mu_k)p(\mu_k|\theta_{\alpha}) p_{\alpha} }{\int d\mathcal{R} d\Theta dP\, (\prod\limits_{i\in\mathcal{E}} d\mu_i)\, p( \{\mu\}_\mathcal{E}, \mathcal{R}, \Theta, P, \{\data\}_\mathcal{E}) } .
\end{align}
Comparing such posterior probabilities for different subpopulations (e.g., NSBH vs BBH) is the basis of our inference.
This is done most straightforwardly by computing an odds ratio
\begin{equation}\label{eq:odds ratio}
    \mathcal{O}^{\alpha_k}_{\beta_k} = \frac{p(\alpha_k|\{\data\}_\mathcal{E})}{p(\beta_k|\{\data\}_\mathcal{E})}
\end{equation}
between two possible classifications $\alpha$ and $\beta$. 

We can gain a bit more intuition for the posterior probability~\eqref{eq:full posterior prob} by rewriting it as
\begin{equation} \label{eq: posterior prob rewrite}
    p(\alpha_k|\{\data\}_\mathcal{E}) = q_{\alpha_k|\{\data\}_{\mathcal{E} \setminus k}} \int d\mu_k\, p(\data_k|\mu_k) q(\mu_k|\{\data\}_{\mathcal{E} \setminus k},\alpha_k) ,
\end{equation}
where $\{\data\}_{\mathcal{E} \setminus k}$ is the set of observed data from the $N-1$ systems not including the $k^\mathrm{th}$ event,
\begin{equation}
   q(\mu_k|\{\data\}_{\mathcal{E} \setminus k},\alpha_k) = \frac{\int d\mathcal{R} d\Theta dP\, p( \mathcal{R},\Theta, P) \mathcal{R}^N e^{-\mathcal{R}\beta(\Theta,P)} \left( \prod\limits_{i \neq k}  d\mu_i\, p(\data_i|\mu_i) p(\mu_i|\Theta, P) \right) p(\mu_k|\theta_\alpha) p_{\alpha} }{q_{\alpha_k|\{\data\}_{\mathcal{E} \setminus k}}}
\end{equation}
and
\begin{equation}
    q_{\alpha_k|\{\data\}_{\mathcal{E} \setminus k}} = \frac{\int d\mathcal{R} d\Theta dP\, p(\mathcal{R}, \Theta, P) \mathcal{R}^N e^{-\mathcal{R}\beta(\Theta,P)} \left(\prod\limits_{i\neq k} d\mu_i\, p(\data_i|\mu_i)p(\mu_i|\Theta, P) \right) p_{\alpha}}{\int d\mathcal{R} d\Theta dP\, p(\mathcal{R}, \Theta, P) \mathcal{R}^N e^{-\mathcal{R}\beta(\Theta,P)} \prod\limits_{j\neq k} d\mu_j\, p(\data_j|\mu_j)p(\mu_j|\Theta, P)} .
\end{equation}
\end{widetext}
This has the natural interpretation of a properly normalized population-informed prior distribution $q(\mu_k|\{\data\}_{\mathcal{E} \setminus k},\alpha_k)$ for the systems in the subpopulation $\alpha_k$ and a population-informed prior probability $q_{\alpha_k|\{\data\}_{\mathcal{E} \setminus k}}$ of belonging to the subpopulation $\alpha_k$.
Indeed, the population-informed mass distribution $q(\mu_k|\{\data\}_{\mathcal{E} \setminus k},\alpha_k)$ is just the single-event parameter prior $p(\mu_k|\theta_{\alpha_k})$ marginalized over the uncertainty in the population parameters conditioned on data from the $N-1$ other detections and the fact that we have detected $N$ systems in total.
In this sense, the hierarchical population inference automatically determines the correct priors for the single-event analyses by marginalizing over population-level uncertainty.
Thus, if we have enough events in our catalog so that the population is well measured, we can take the measured population at face value when interpreting single events.

Alternatively, from Eq.~\eqref{eq: posterior prob rewrite}, we can express the odds ratio as
\begin{equation}\label{eq:decomposed odds}
    \mathcal{O}^\mathrm{\alpha_k}_\mathrm{\beta_k} = \mathcal{B}^\mathrm{\alpha_k}_\mathrm{\beta_k} \left( \frac{ q_{\alpha_k|\{\data\}_{\mathcal{E} \setminus k}} }{ q_{\beta_k|\{\data\}_{\mathcal{E} \setminus k}} } \right) ,
\end{equation}
where
\begin{equation}
    \mathcal{B}^\mathrm{\alpha_k}_\mathrm{\beta_k} = \frac{\int d\mu_k\, p(\data_k|\mu_k) q(\mu_k|\alpha_k, \{\data\}_{\mathcal{E} \setminus k})}{\int d\mu_k\, p(\data_k|\mu_k) q(\mu_k|\beta_k, \{\data\}_{\mathcal{E} \setminus k})}
\end{equation}
is a Bayes factor given population-informed mass priors and $(q_{\alpha_k|\{\data\}_{\mathcal{E} \setminus k}} / q_{\beta_k|\{\data\}_{\mathcal{E} \setminus k}})$ serves as the population-informed prior odds.

\subsection{Classification metric: Odds ratio vs Bayes factor}
\label{sec:odds vs bayes}

While one might naturally attempt to distinguish between possible types of compact binaries based on the Bayes factor $\mathcal{B}^\alpha_\beta$ in an attempt to minimize the impact of the prior odds in Eq.~\eqref{eq:decomposed odds}, it turns out that $\mathcal{O}^\alpha_\beta$ is actually less sensitive to prior assumptions.
Consider the following, in which we omit the label $k$ for the compact object of interest.
Assuming a fixed population so that $q(\mu|\{\data\}_{\mathcal{E} \setminus k},\alpha) = p(\mu|\theta_\alpha)$, and focusing for the moment on a comparison between the NSBH and BBH subpopulations, we have
\begin{equation}\label{eq:fixed pop bayes}
    \mathcal{B}^\mathrm{NSBH}_\mathrm{BBH} = \frac{\int d\mu\, p(\data|\mu) p(\mu|\theta_\mathrm{NSBH})}{\int d\mu\, p(\data|\mu) p(\mu|\theta_\mathrm{BBH})} .
\end{equation}
Because BBHs span a larger mass range than NSBHs (the secondary is limited to $m_2 \leq \Mmax$), $p(\mu|\mathrm{BBH})$ brings along a relatively large Occam factor that can severely penalize the BBH model if the likelihood has support over only a small mass range.
In other words, \emph{$\mathcal{B}^\mathrm{NSBH}_\mathrm{BBH}$ can be quite sensitive to the high-mass behavior of $p(\mu|\theta_\mathrm{BBH})$ even though the likelihood is vanishingly small at such large masses}.
We can see this more explicitly by including the normalization of the priors
\begin{equation}\label{eq:fixed pop bayes with norm}
    \mathcal{B}^\mathrm{NSBH}_\mathrm{BBH} = \frac{\int d\mu\, p(\data|\mu) d\mathcal{N}_\mathrm{NSBH}/d\mu}{\int d\mu\, p(\data|\mu) d\mathcal{N}_\mathrm{BBH}/d\mu} \left( \frac{\int d\mu\, d\mathcal{N}_\mathrm{BBH}/d\mu}{\int d\mu\, d\mathcal{N}_\mathrm{NSBH}/d\mu} \right) .
\end{equation}
If our population model predicts approximately equal numbers of NSBHs and BBHs within the likelihood's support, then the first ratio is of order unity.
However, the second ratio can confound this, as the BBH mass distribution may extend to much higher masses and there may simply be more BBHs than NSBHs in the Universe.
If this is the case, then we would infer that $\mathcal{B}^\mathrm{NSBH}_\mathrm{BBH} \gg 1$ based primarily on our knowledge of the high-secondary-mass BH distribution, which should be irrelevant when classifying compact binaries with low-mass secondaries.

We also note that, were we to assume approximately equal numbers of NSBHs and BBHs in the Universe, and therefore equal prior odds, this would imply either an extremely steeply falling BBH number density with increasing mass or a steep feature in the total mass distribution at or below \Mmax.
In general, it is not possible to assign arbitrary prior odds to the different components of a mixture model while self-consistently fixing the shapes of both the subpopulations and overall mass distributions.

In contrast to $\mathcal{B}^{\rm NSBH}_{\rm BBH}$, the odds ratio becomes
\begin{equation}
    \mathcal{O}^\mathrm{NSBH}_\mathrm{BBH} = \frac{\int d\mu\, p(\data|\mu) d\mathcal{N}_\mathrm{NSBH}/d\mu}{\int d\mu\, p(\data|\mu) d\mathcal{N}_\mathrm{BBH}/d\mu} ,
\end{equation}
which only depends on the number densities within the likelihood's support.
That is to say, \emph{$\mathcal{O}^\mathrm{NSBH}_\mathrm{BBH}$ only depends on our knowledge of the mass distributions for masses similar to this event's}.
While this requires us to specify the number density of compact objects within particular mass ranges, rather than just a normalized distribution, this is a more physically relevant representation anyway.
As such, we use $\mathcal{O}^\alpha_\beta$ rather than $\mathcal{B}^\alpha_\beta$ as our classification metric throughout the rest of this work.

\section{Population models}
\label{sec:sensitivity to population parameters}

So far, we have described a general classification scheme that places no restrictions on the subpopulations.
Indeed, Eq.~\eqref{eq:full posterior prob} allows for classification while simultaneously accounting for uncertainty from the population inference.
However, as even the functional form of the overall mass distribution is not yet tightly constrained~\citep{GWTC1RatesAndPopulations, MatterMatters}, we are faced with several different sources of uncertainty.
Section~\ref{sec:massdistribs} discusses uncertainty in the overall mass distribution.
Section~\ref{sec:mmax} considers our uncertainty in the NS \EOS, \Mtov, and \Mmax.
Section~\ref{sec:overlap} discusses the uncertainty in subpopulations, particularly whether the maximum NS produced in nature is limited by the formation channel rather than the EOS ($\Mform < \Mmax$) and whether BHs and NSs exist within the same mass range.

\subsection{Overall mass distribution}
\label{sec:massdistribs}

Several authors have studied the astrophysical distribution of both NS and BH masses (see, e.g., \citet{AntoniadisTauris2016, Farrow2019, Chatziioannou2020, FarrChatziioannou2020, Alsing2018, Fishbach_2017, GWTC1RatesAndPopulations, FishbachFarrHolz, PickyPartners} as well as the reviews in \citet{2018arXiv180605820M} and \citet{Postnov2014}).
Specific distributions for NSs or BHs may be motivated theoretically through population synthesis calculations or empirically through observational surveys, and may have rather complex shapes.
Nevertheless, our knowledge of the overall rate density of low-mass compact objects $d\mathcal{N}_\mathrm{tot}/d\mu$ may quickly become more precise than our knowledge of the subpopulation rate densities $d\mathcal{N}_\alpha/d\mu$~\citep{Wysocki2020}.
This is because only relatively loud GW signals will carry enough tidal information to clearly signal the presence of a NS~\citep{Lackey2015, Landry2020, ChenJohnsonMcDaniel2020, Fasano2020}, and we are likely to have many more quiet detections than loud detections.
Furthermore, depending on the distributions of binaries containing NSs, most of the loud BNS and NSBH detections may involve NSs with $m \ll \Mmax$, thereby providing little information about the upper reaches of the NS mass distribution.

We also note that the selection effects $\beta(\Theta, P)$ in our posteriors (Eqs.~\ref{eq: joint} and~\ref{eq: masspost}) only depend on the total distribution $p(\mu|\Theta, P)$ and not on the individual subpopulation distributions $p(\mu|\theta_\alpha)$.
Therefore, assuming precise knowledge of the overall mass distribution removes any dependence on selection effects from our inference when we marginalize over \Mmax.
As such, we consider several fixed overall mass distributions, rather than specifying distributions for, e.g., each of the BNS, NSBH, and BBH subpopulations.
Specifically, we assume a few basic mass distributions for individual compact objects, all of which take the form
\begin{align}
    p(m) & \propto \mathrm{H}(m \leq M_\mathrm{brk}) m^\alpha + \Delta \times \mathrm{H}(M_\mathrm{brk} < m) m^\alpha \label{eq:p_Delta} ,
\end{align}
where, $H(\cdot)$ is the Heaviside function.
We further assume the mass distribution to be independent of spin and other source properties.
Our choices for the parameters $\alpha$, $\Delta$ and $M_{\rm brk}$ are listed in Table~\ref{tab:models}.
The value of $\alpha$ is based on the inferred exponent from higher-mass BBH mergers during O1 and O2 \citep{GWTC1RatesAndPopulations}.
Eq.~\eqref{eq:p_Delta} is also motivated by the expectation for a low-mass gap between NSs and BHs \citep{2010ApJ...725.1918O, Bailyn_1998, Farr_2011}, which we model as a sharp decrease in the overall mass distribution at $M_\mathrm{brk}$.
Our specific choices for $M_\mathrm{brk}$ and $\Delta$ are \textit{ad hoc}; they are meant to simulate a sharp feature in $d\mathcal{N}_\mathrm{tot}/dm$ near the median of our current uncertainty in \Mtov~and proposed upper limits from EM observations of AT 2017gfo \citep{PhysRevD.100.023015,GW170817-ModelSelection}, although there is some disagreement about the exact value of that upper limit \citep[e.g.,][]{Margalit_2017, Rezzolla_2018, Ai_2020}.
However, $M_\mathrm{brk}$ may or may not be related to \Mmax~(or even \Mform), as this would depend on the formation channel \citep[e.g.,][]{Ertl_2020}; there could be BHs with $\Mmax < m \leq M_\mathrm{brk}$.
What's more, incorrectly assuming $M_\mathrm{brk}\approx\Mmax$ could lead to biases in \EOS~constraints \citep{Miller2019b, Landry2020}.
In this section, we fix $M_\mathrm{brk}$ but do not allow this choice to influence our uncertainty in \Mtov~or \Mmax.

Following \citet{PickyPartners} and \citet{MatterMatters}, we construct joint distributions for $m_1$ and $m_2$ that are proportional to these single-component mass distributions and a pairing function, such that
\begin{equation}
    p(m_1, m_2) \propto p(m_1) p(m_2) q^\beta \mathrm{H}(m_1 \geq m_2) \label{eq:p_joint} ,
\end{equation}
where $q=m_2/m_1$ and $\beta=4$.
We also consider random pairing, which corresponds to $\beta = 0$.
We also explicitly impose our convention that $m_1 \geq m_2$.
As different choices of $\alpha$, $\beta$, $\Delta$, and $M_\mathrm{brk}$ change our quantitative results, we report results with a few example distributions in Tables~\ref{tab:GW190425} and~\ref{tab:GW190814} to give a sense of the variability.
We do not claim that these choices represent all possible mass distributions, but instead that they demonstrate the general conclusions for GW190425's $m_1$ and GW190814's $m_2$.

\begin{table}
    \caption{
        Single-component mass distributions assumed in this analysis, all realizations of Eq.~\eqref{eq:p_Delta} with different parameters.
        We note that when $\Delta=1$, the distribution does not depend on value of $M_\mathrm{brk}$.
        Specific combinations of these distributions are considered in Tables~\ref{tab:GW190425} and~\ref{tab:GW190814} to quantify the sensitivity to the unknown distribution of low-mass compact objects.
    }
    \label{tab:models}
    \begin{tabular}{cccc}
        \hline
                              & $\alpha$ & $\Delta$ & $M_\mathrm{brk}$ \\
        \hline
        \hline
        $p_0(m)$      & 0.0 & 1.0 & arbitrary \\
        $p_\mathrm{PL}(m)$ & -1.3 & 1.0 & arbitrary \\
        $p_\mathrm{BRK}(m)$ & -1.3 & 0.1 & $2.3\,M_\odot$ \\
        \hline
    \end{tabular}
\end{table}

\subsection{Maximum neutron star mass}
\label{sec:mmax}

Because the NS mass distribution truncates at (or below) \Mmax, our uncertainty in \Mmax~is directly tied to our ability to confidently identify individual objects as BHs rather than NSs.
\emph{We therefore explore how our knowledge that a single subpopulation must truncate at a particular mass scale}, which may be determined outside our population analysis and specified as a prior in Eq.~\eqref{eq:full posterior prob}, \emph{affects our ability to distinguish between BHs and NSs.} Previous studies assumed an exact, fixed boundary between those two subpopulations, but we instead use current knowledge from theoretical and empirical studies of NSs. We assume our uncertainty in \Mmax~is uncorrelated with the mass of the compact object of interest, although this may not truly be the case if we simultaneously infer both the \EOS~and the mass distribution~\citep{Wysocki2020}.

There are several estimates for \Mmax~and \Mtov~in the literature, derived from different astrophysical observables.
For instance, our knowledge of \Mtov~is informed by observations of massive pulsars~\citep{Cromartie:2019kug, Antoniadis:2013pzd}, GWs from GW170817 and GW190425, and X-ray timing observations of PSR J0030+0451~\citep{Riley2019, Miller2019, Raaijmakers2019} as well as the EM counterparts from GW170817~\citep{Dietrich2020, GW170817-ModelSelection, Margalit_2017, Rezzolla_2018, PhysRevD.100.023015}.
Similarly, studies of the mass distribution of Galactic NSs~\citep{FarrChatziioannou2020, Alsing2018} constrain \Mform, a lower limit for \Mmax.
Moreover, the relation between $\Mtov$ and \Mmax~for rotating NSs has been investigated through numerical studies of rapidly spinning relativistic stars \citep[e.g.,][]{1994ApJ...424..823C,PhysRevLett.111.131101,Rezzolla_2018}.

We explore a few proposed \Mmax~distributions to characterize our inference's sensitivity to this uncertainty.
Specifically, we use the inferred posterior distribution for the maximum gravitational mass of a non-rotating NS, \Mtov, from \citet{Landry2020} based on a nonparametric analysis of massive pulsar, GW, and X-ray timing data.
We compare this to an analysis of GW170817 assuming a spectral \EOS~parametrization \citep{Carney2018} and $\Mtov\geq1.97\,M_\odot$ \citep{GW170817-EOS}, although the spectral parametrization may introduce model systematic errors~\citep{Tan2020}, particularly in comparison to the nonparametric \EOS~inference of \citet{Landry2020}.

We also study an empirical fit to observed galactic NSs \citep{FarrChatziioannou2020} that includes a \Mmax~parameter.
For the \citet{Landry2020} $\Mtov$ prediction, we additionally investigate the effect of upper limits estimated from numerical simulations of ejected mass and kilonova luminosity coupled with observations of AT 2017gfo \citep{PhysRevD.100.023015,GW170817-ModelSelection}, which suggest $\Mtov \lesssim 2.3\,\Msolar$ and $\Mmax \lesssim 2.7\,\Msolar$.

\subsection{Overlap of neutron star and black hole mass distributions}
\label{sec:overlap}

Eq.~\eqref{eq:full posterior prob} suggests that we should directly marginalize over our uncertainty in \Mmax.
This is straightforward, but requires knowledge of the individual subpopulation's rate densities.
While we can confidently state that NS cannot exist with $m > \Mmax$ (by definition), and there are reasons to believe that BHs do not exist below \Mmax~\citep[e.g.,][]{Belczynski_2012, Fryer_2001}, we cannot be certain that $d\mathcal{N}_\mathrm{BBH}/d\mu$ identically vanishes below \Mmax.
For example, primordial BHs could form in this mass range, pair, and merge within a Hubble time~\citep{CarrHawking1974,Meszaros1974}.
Given the fact that we will likely measure only $d\mathcal{N}_\mathrm{tot}/d\mu$ directly and not the rate densities of separate subpopulations, we restrict our study to mass distributions with no overlap between NSs and BHs.
This amounts to asking whether individual objects are above or below \Mmax, or equivalently maximizing the posterior probability that any individual object is a NS while fixing $d\mathcal{N}_\mathrm{tot}/d\mu$.
That is, with the current uncertain state of our population knowledge, we can confidently rule out NSs, but we cannot confirm their presence without more detailed knowledge of the subpopulations.

Similarly, astrophysical formation channels may limit NSs to masses $m\leq \Mform \leq \Mmax$.
Assuming $\Mform=\Mmax$ also maximizes the probability that an object is a NS while fixing the overall mass distribution.

If we assume a known total mass distribution and that compact binaries are composed of only NSs and BHs, it implies the following.
We denote the rate density of systems where the object in question is a NS as $d\mathcal{N}_\mathrm{NS}/d\mu$ (either a BNS or NSBH depending on the system) and the case where the object is a BH as $d\mathcal{N}_\mathrm{BH}/d\mu$ (similarly, either NSBH or BBH).
Then
\begin{gather}
    \frac{d\mathcal{N}_\mathrm{NS}}{d\mu} \leq \int d\Mmax\, p(\Mmax) \frac{d\mathcal{N}_\mathrm{tot}}{d\mu}\mathrm{H}\left(m \leq \Mmax\right) , \\
    \frac{d\mathcal{N}_\mathrm{BH}}{d\mu} \equiv \frac{d\mathcal{N}_\mathrm{tot}}{d\mu} - \frac{d\mathcal{N}_\mathrm{NS}}{d\mu} ,
\end{gather}
and therefore
\begin{equation} \label{eq:nooverlap}
    \mathcal{O}^\mathrm{NS}_\mathrm{BH} \leq \frac{P(m \leq \Mmax)}{1 - P(m \leq \Mmax)}
\end{equation}
where
\begin{align}\label{eq:this one}
    P(m & \leq \Mmax) = \nonumber \\ & \int d\Mmax\, p(\Mmax) \int dm\, p(m|\data) \mathrm{H}(m \leq \Mmax) .
\end{align}
Although Eq.~\eqref{eq:this one} explicitly calls out the single object's mass as the variable of interest, we remind the reader that \Mmax~depends on \Mtov~and the object's spin, and $\mathrm{H}(m \leq \Mmax)$ should be thought of as a condition in the multi-dimensional space spanned by an individual object's mass, spin, and \Mtov.

\begin{figure}
    \begin{center}
        \includegraphics[width=\columnwidth, clip=True, trim=0.1cm 0.3cm 0.1cm 0.1cm]{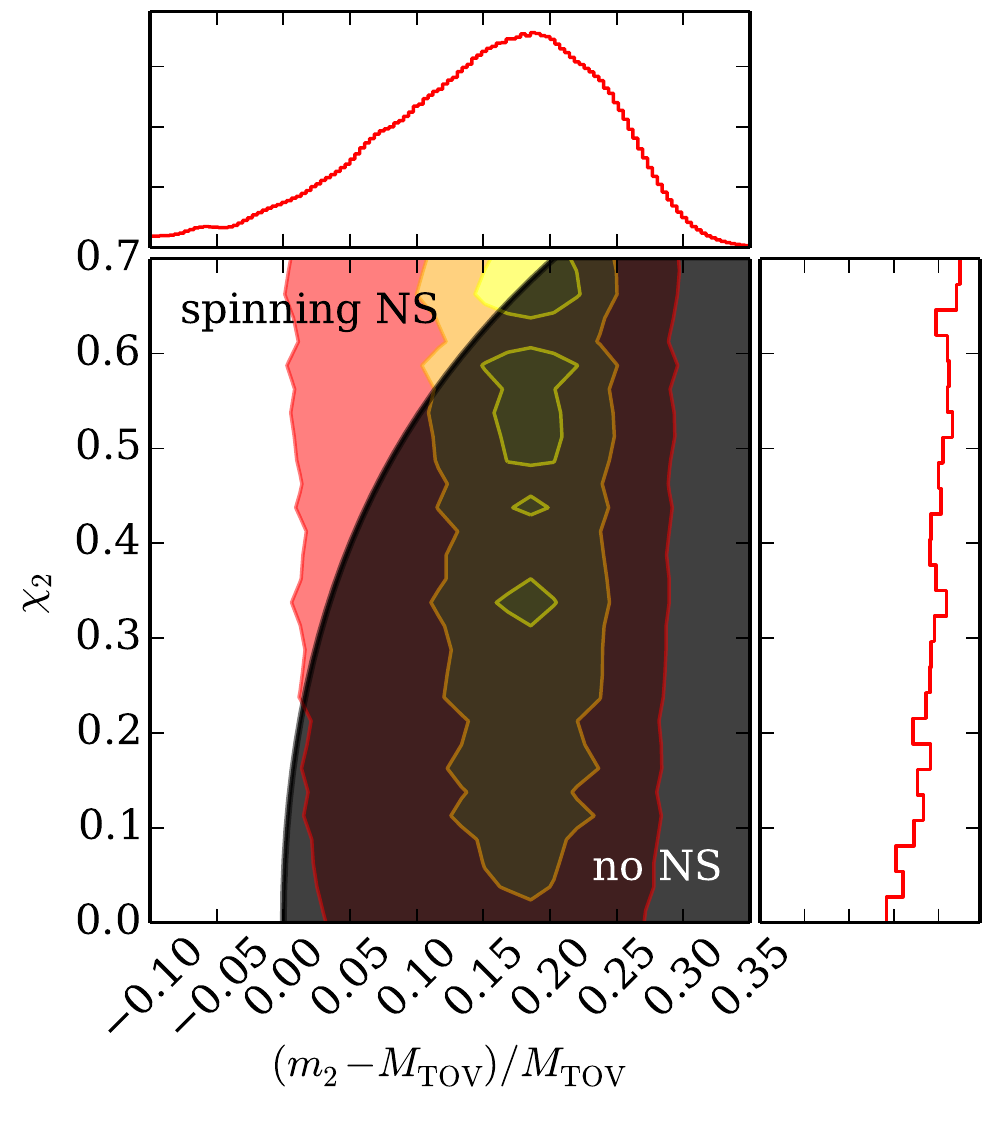}
    \end{center}
    \caption{
        Joint posterior distribution for the relative difference between GW190814's $m_2$ and \Mtov~along with the object's dimensionless spin $\chi_2$ (shown in red).
        The right panel demonstrates that the spin is essentially unconstrained (the posterior is nearly identical to the prior) and uncorrelated with $m_2$.
        The main panel is divided into regions compatible (unshaded) and incompatible (shaded) with stable NSs, based on mass and spin.
        The separatrix is taken from the universal relation of \citet{Breu2016} between spin and the rotationally supported maximum NS mass.
        The bulk of the posterior probability lies within the NS exclusion zone, except if we allow for extreme values of the secondary's spin.
    }
    \label{fig:GW190814 m2 vs chi2 correlations}
\end{figure}

Although the maximum spin a NS can attain depends on the \EOS, several studies have found that the dimensionless spin will be limited to $\chi \lesssim 0.7$ \citep[e.g.,][]{Essick2020, 1994ApJ...424..823C, 1995A&A...296..745H, Lattimer2001}.
At the same time, maximally spinning NSs without differential rotation are thought to reach masses between 1.2--1.3$\Mtov$ \citep{Breu2016, Rezzolla_2018}, although some estimates can be larger \citep{PhysRevLett.111.131101}.
While universal relations exist that relate $\chi$ and \Mmax~\citep[e.g.,][]{Breu2016}, which should mitigate the effects of our uncertainty in the \EOS, these were constructed by considering only {\EOS}s~without strong phase transitions.
As such, we primarily investigate limiting cases where either $\Mmax = \Mtov$ or $\Mmax = 1.3\Mtov$, regardless of the object's spin.
This bounds how much the scaling between $\Mmax$ and $\chi$ could affect our analysis, although Fig.~\ref{fig:GW190814 m2 vs chi2 correlations} sketches the higher-dimensional inference for GW190814's $m_2$ with more precise knowledge of $\Mmax(\Mtov, \chi)$, similar to what is discussed in \citet{Most2020}.
We expect $d\Mmax/d\chi,\ d^2\Mmax/d\chi^2 > 0$ because, as the star oblates under the influence of its own spin, the centrifugal force at the surface will increase while the surface gravity simultaneously decreases.
Therefore, we should expect a convex separatrix between NSs and BHs, such as the universal relation from~\citet{Breu2016} shown in Fig.~\ref{fig:GW190814 m2 vs chi2 correlations}.
The convexity of the separatrix requires more extreme values of the spin to support NSs with masses significnatly above \Mtov, as compared to a spin-independent scaling like $\Mmax = 1.3\Mtov$.

\section{Case Studies}
\label{sec:examples}

We calculate $P(m \leq \Mmax)$ for GW190425's primary component (Table~\ref{tab:GW190425}) and GW190814's secondary component (Table~\ref{tab:GW190814}) with a few choices of population models (Table~\ref{tab:models}) along with the different estimates of \Mmax~(Section~\ref{sec:mmax}).
We assume no overlap between the NS and BH mass distributions in order to maximize $P(m \leq \Mmax)$.

\subsection{GW190425}
\label{sec:190425}

\begin{table*}
    \caption{
        Estimates of $P(m_1\leq\Mmax)$ for GW190425, the posterior probability that the primary component's mass is compatible with a NS, under different assumptions about the compact object population (columns) and the maximum NS mass $\Mmax$ (rows).
        We report means $\pm$ standard deviations from Monte-Carlo integration.
        The component mass priors are assumed to be either a uniform distribution $p_0(m)$, a power-law distribution $p_{\textsc{pl}}(m)$, or a power-law distribution with a mass gap $p_{\textsc{brk}}(m)$ as defined by Eq.~\eqref{eq:p_Delta} with the parameter choices listed in Table~\ref{tab:models}.
        The pairing of compact objects is assumed to be random or mass-ratio dependent, such that the population prior defined in Eq.~\eqref{eq:p_joint} takes the form $p(m_1,m_2) = p(m_1)p(m_2)$ or $p(m_1,m_2) = p(m_1)p(m_2)q^4$, respectively.
        The various $M_{\rm max}$ distributions account for different constraints on the maximum NS mass and different assumptions about the primary's spin.
        The first three rows assume negligible spin, such that $\Mmax=\Mtov$, and adopt the \Mtov~posterior from \citet{Landry2020} (LEC) or \citet{GW170817-EOS} (LVC).
        The first row additionally accounts for an upper bound of $\Mtov \leq 2.3\,M_\odot$ motivated by EM observations of AT 2017gfo \citep{PhysRevD.100.023015,GW170817-ModelSelection}.
        The fourth row uses the NS population-based estimate of $\Mmax$ from \citet{FarrChatziioannou2020} (FC), allowing for spin.
        The last row assumes rotation at the NS breakup frequency, such that $\Mmax=1.3\Mtov$, while accounting for the upper bound of $\Mmax \leq 2.7\,M_\odot$ on the maximum mass of a rotating NS from AT 2017gfo.
    }
    \label{tab:GW190425}
    \begin{center}
        \begin{tabular}{ccccccc}
        \hline
        \multicolumn{2}{c}{\multirow{2}{*}{$\Mmax$}} & \multicolumn{5}{c}{population prior $p(m_1, m_2)$} \\
        \cline{3-7}
        & & $p_0(m_1)p_0(m_2)$ & $p_\mathrm{PL}(m_1)p_0(m_2)$ & $p_\mathrm{PL}(m_1) p_\mathrm{PL}(m_2) q^4$ & $p_\mathrm{BRK}(m_1)p_0(m_2)$ & $p_\mathrm{BRK}(m_1) p_\mathrm{BRK}(m_2) q^4$ \\
        \hline
        \hline
        \multirow{1}{*}{LEC}
            & $\Mtov \leq 2.3\,\Msolar$ & \result{61.67 \pm 0.75 \%} & \result{68.99 \pm 0.82 \%} & \result{91.9 \pm 1.0 \%} & \result{86.09 \pm 0.97 \%} & \result{95.7 \pm 1.0 \%} \\
        \multirow{1}{*}{LEC}
            & $\Mtov$ & \result{68.58 \pm 0.69 \%} & \result{74.88 \pm 0.73 \%} & \result{93.82 \pm 0.87 \%} & \result{90.09 \pm 0.84 \%} & \result{97.03 \pm 0.89 \%} \\
        \multirow{1}{*}{LVC}
            & $\Mtov$ & \result{62.0 \pm 1.3 \%} & \result{68.8 \pm 1.4 \%} & \result{90.9 \pm 1.8 \%} & \result{83.6 \pm 1.7 \%} & \result{94.2 \pm 1.9 \%} \\
        \multirow{1}{*}{FC}
            & $\Mmax$ & \result{72.9 \pm 1.2 \%} & \result{77.9 \pm 1.2 \%} & \result{93.6 \pm 1.4 \%} & \result{88.7 \pm 1.4 \%} & \result{96.0 \pm 1.5 \%} \\
        \multirow{1}{*}{LEC}
            & $1.3 \Mtov \leq 2.7\,\Msolar$ & \result{96.5 \pm 1.5 \%} & \result{97.6 \pm 1.5 \%} & \result{99.7 \pm 1.5 \%} & \result{99.7 \pm 1.5 \%} & \result{99.9 \pm 1.5 \%} \\
        \hline
    \end{tabular}
    \end{center}
\end{table*}

GW190425 is the second BNS candidate detected by LIGO and Virgo.
The secondary component of GW190425 has a mass constrained within 1.12--1.68\,\Msolar.
Given our assumption of non-overlapping NS and BH mass distributions, this immediately identifies the secondary as a NS, since we know $\Mmax \gtrsim 2\,\Msolar$~\citep{Cromartie:2019kug}.
The primary component of 1.61--2.52\,\Msolar~could, in principle, be either a NS or BH.
Hence, we compute the odds ratio $\mathcal{O}^{\rm BNS}_{\rm NSBH}$ according to Eq.~\eqref{eq:nooverlap}, which depends only on $P(m_1 \leq \Mmax)$, to classify the primary component.
We evaluate $P(m_1 \leq \Mmax)$ via Monte-Carlo integrals over reweighted public posterior samples for GW190425 \citep{GW19425samples}, listing the results in Table~\ref{tab:GW190425} given a variety of assumptions. 

We find that the uncertainty in the assumed overall mass distribution leads to variation at least as large, if not larger, than uncertainty in the \Mmax~distribution.
This is likely because GW190425 has a relatively low S/N, implying that its likelihood is not strongly peaked, and the posterior is sensitive to the assumed prior.
Also, much of GW190425's $m_1$ posterior is below the smallest \Mmax~allowed by any of the distributions we consider, and therefore the uncertainty in \Mmax~does not matter for a sizeable fraction of the possible $m_1$ values.
Generally, we find that using any reasonable population prior that is not flat in both $m_1$ and $m_2$ introduces a preference for both smaller component masses and mass ratios close to unity.
Both these effects tend to concentrate the $m_1$ posterior at lower values, thereby raising our confidence that it is below \Mmax.
Indeed, we find that it is quite likely that $m_1$ was a non-spinning NS, and there is only a \result{\lesssim 1\%} chance that $m_1$ was so large as to rule out maximally spinning NSs.
While this does not prove that either of GW190425's components were NSs, it reiterates that the system is completely consistent with a BNS.

\citet{Foley2020} propose a few specific astrophysically-motivated formation scenarios that, contrary to our assumptions, tend to favor more asymmetric mass ratios.
They show that GW190425 is consistent with the coalescence of a low-mass BH and a NS, but do not attempt to quantify the posterior odds for that hypothesis.
Similarly, \citet{Han_2020} explore GW190425's consistency with a NSBH merger.
Our analysis confirms that this interpretation is compatible with the data, but it suggests the event is more likely to have been a BNS merger.

\subsection{GW190814}
\label{sec:190814}

GW190814 is an unequal mass ratio coalescence detected by LIGO and Virgo and initially announced as a NSBH candidate.
An abbreviated version of this analysis in \citet{GW190814} raised the strong possibility that its secondary ($m_2 \approx 2.6\,\Msolar$) was a BH rather than a NS~\citep{GW190814}.
The nature of its $23\,\Msolar$ primary BH is not in doubt, however.
We therefore compute $P(m_2 \leq \Mmax)$ to estimate $\mathcal{O}^{\rm NSBH}_{\rm BBH}$.

Table~\ref{tab:GW190814} reports $P(m_2\leq\Mmax)$ for GW190814 using publicly available posterior samples~\citep{GW190814samples}.
In this case, the effect of the population prior is negligible and instead most of our systematic uncertainty comes from the \Mmax~distribution.
This is likely because GW190814 has a higher S/N than GW190425, and its asymmetric mass ratio makes higher-order modes in the GW signal more important.
The presence of detectable higher-order modes can break degeneracies within the GW waveform and improve the likelihood's constraints on $q$.
As such, we have a much more precise constraint on $m_2$ that is less sensitive to our assumptions about the underlying mass distribution.
This includes the presence of sharp features within $p(m|\Theta, P)$.
In particular, if we include a steep mass-gap feature, it must be several orders of magnitude deep ($\Delta \lesssim 10^{-2}$) in order to select only the tail of the $m_2$ distribution over the much larger likelihoods at higher masses.
Even if this is the case, that tail does not extend significantly below $2.3\,M_\odot$ (the smallest $m_2$ value from the $\sim 2800$ public samples is $2.3\,M_\odot$), which is still above a significant fraction of the \Mmax~distributions and therefore corresponds to relatively small $P(m_2\leq \Mmax)$.
We explore the assumption of perfect mass gaps, the limit $\Delta \rightarrow 0$, in Section~\ref{sec:mass gap}.

We also note relatively large differences when assuming \Mmax~distributions based on constraints placed on the NS \EOS~from the existence of massive pulsars, GWs from coalescences known to contain at least one NS, and X-ray timing of rapidly spinning pulsars \citep{Landry2020, GW170817-EOS} compared to empirical fits that are not constrained by nuclear physics \citep{FarrChatziioannou2020}.
This is because the empirical fit to the NS mass distribution has a much larger tail to high \Mmax~compared to the \Mtov~distributions based on the inferred \EOS.
We have checked that if we impose the equivalent of an upper limit of $\Mtov \leq 2.3\,M_\odot$ based on EM observations of AT 2017gfo \citep{PhysRevD.100.023015,GW170817-ModelSelection} on the empirical \Mmax~distributions, there is much better agreement between the different approaches.

Interestingly, essentially all of the $m_2$ posterior falls between \Mtov~and 1.3\Mtov~regardless of the systematic uncertainty in \Mmax.
This means that it is unlikely that $m_2$ was a non-spinning NS, but $m_2$ remains consistent with a spinning NS.
Fig.~\ref{fig:GW190814 m2 vs chi2 correlations} shows that the secondary's spin is nearly unconstrained, and therefore any finer grained inference about whether $m_2$ and $\chi_2$ could correspond to a NS will depend strongly on the assumed spin prior.
As such, we do not attempt to quantify this, but note that the separatrix between spinning NSs and BHs in the $m$--$\chi$ plane will be convex, like the universal relation reported in ~\cite{Breu2016}.
This could suggest that GW190814 is more consistent with a BBH coalescence than a NSBH, particularly if astrophysical NSs can only form with relatively small spins.
Indeed, Galactice NSs in binaries that will merge within a Hubble time have $\chi\leq0.05$~\citep{GW170817-Props, Essick2020} and the fastest known pulsar spin corresponds to $\chi\sim 0.35$ \citep{Hessels2006, Essick2020}, which may be small enough that $\Mmax\approx\Mtov$.

\citet{Most2020} take this line of reasoning further.
Utilizing approximate universal relations for $\Mmax(\Mtov, \chi)$ from \citep{Breu2016}, they bound both $\Mtov$ and $\chi_2$ from below, finding $\Mtov > 2.08 \pm 0.04\, M_\odot$ and $\chi_2 \gtrsim 0.49$, as fast or faster than the fastest known pulsar \citep{Hessels2006}.
Their findings are consistent with our results, as we show that $m_2$ is unlikely to have been a slowly spinning NS, but they additionally assume $m_2$ must have been a NS at some point.
They reason that the collapse from a NS to a BH must have occured along the universal reation for $\Mmax(\Mtov,\chi)$ and correspondingly impose a tight prior on the objects spin as a function of mass.
However, as we do not make these assumptions, we cannot place similar bounds.

\begin{table*}
    \caption{
        Estimates of $P(m_2\leq\Mmax)$ for GW190814, the posterior probability that the secondary component's mass is compatible with a NS, under different assumptions about the compact object population (columns) and the maximum NS mass $\Mmax$ (rows).
        We report means $\pm$ standard deviations from Monte-Carlo integration.
        The assumed population models and $\Mmax$ distributions are the same as in Table~\ref{tab:GW190425}.
        Because the smallest $m_2$ sample in the GW190814 posterior data is $\sim 2.3\,M_\odot$, we only provide approximate upper limits in the first row, where the maximum mass distribution is truncated to $\Mmax\leq2.3\,M_\odot$.
        We note that the first entry in the third row, which assumes \Mmax=\Mtov~from \cite{GW170817-EOS} (LVC), recovers the result presented in \citet{GW190814}.
    }
    \label{tab:GW190814}
    \begin{center}
        \begin{tabular}{ccccccc}
        \hline
        \multicolumn{2}{c}{\multirow{2}{*}{$\Mmax$}} & \multicolumn{5}{c}{population prior $p(m_1, m_2)$} \\
        \cline{3-7}
        & & $p_0(m_1)p_0(m_2)$ & $p_\mathrm{PL}(m_1)p_0(m_2)$ & $p_\mathrm{PL}(m_1) p_\mathrm{PL}(m_2) q^4$ & $p_\mathrm{BRK}(m_1)p_0(m_2)$ & $p_\mathrm{BRK}(m_1) p_\mathrm{BRK}(m_2) q^4$ \\
        \hline
        \hline
        \multirow{1}{*}{LEC}
            & $\Mtov \leq 2.3\,\Msolar$ & \result{\leq 0.1\%} & \result{\leq 0.1\%} & \result{\leq 0.1\%} & \result{\leq 0.1\%} & \result{\leq 0.1\%} \\
        \multirow{1}{*}{LEC}
            & $\Mtov$ & \result{5.63 \pm 0.15 \%} & \result{5.55 \pm 0.15 \%} & \result{5.25 \pm 0.15 \%} & \result{5.55 \pm 0.15 \%} & \result{5.25 \pm 0.15 \%} \\
        \multirow{1}{*}{LVC}
            & $\Mtov$ & \result{3.47 \pm 0.32 \%} & \result{3.41 \pm 0.32 \%} & \result{3.18 \pm 0.31 \%} & \result{3.41 \pm 0.32 \%} & \result{3.18 \pm 0.31 \%} \\
        \multirow{1}{*}{FC}
            & $\Mmax$ & \result{29.12 \pm 0.82 \%} & \result{29.02 \pm 0.82 \%} & \result{28.58 \pm 0.82 \%} & \result{29.02 \pm 0.82 \%} & \result{28.58 \pm 0.82 \%} \\
        \multirow{1}{*}{LEC}
            & $1.3\Mtov \leq 2.7\,\Msolar$ & \result{86.9 \pm 1.3 \%} & \result{86.4 \pm 1.3 \%} & \result{83.6 \pm 1.3 \%} & \result{86.4 \pm 1.3 \%} & \result{83.6 \pm 1.3 \%} \\
        \hline
    \end{tabular}
    \end{center}
\end{table*}


\section{Classification-informed single-event and population properties}
\label{sec:mass gap}

Section~\ref{sec:examples} considered the impact of our uncertainty in \Mmax~on our ability to distinguish between NSs and BHs while assuming a fixed overall mass distribution.
Here, we make a different assumption.
We assume an overall mass distribution in such a way that the classification of GW190814 is definite, and then infer how that classification updates our knowledge of the \EOS~and \Mmax.
This shows how definite knowledge about the composition of a system can inform our knowledge of both that system's parameters and population-level parameters.

We assume there is a perfect mass gap starting at \Mmax~and extending to $m\sim 5\,M_\odot$.
We still assume that everything below \Mmax~is a NS, implying that BHs can only exist above $5\,M_\odot$.
Similarly, we assume that the uncertainty in \Mmax~directly translates into uncertainty in the extent of the overall mass distribution below $\sim 5\,M_\odot$.
In contrast to Section~\ref{sec:examples}, this is equivalent to setting $\Delta=0$ and $\Mmax=M_\mathrm{brk}$ in Eq.~\eqref{eq:p_Delta}.

\citet{FishbachFarrHolz} discuss the possibility that noise fluctuations could cause detected systems to look like outliers at first glance when in fact they are entirely consistent with distributions that include sharp cut-offs.
If we assume a perfect mass gap, GW190814's $m_2$ would be an archetypal example of such a system, as it is relatively far down the tail of our uncertainty in \Mmax.
Below, we place a simultaneous posterior over both $m_2$ and \Mmax, although we neglect selection effects associated with the changes in $d\mathcal{N}_\mathrm{tot}/d\mu$ (now assumed to be related to \Mmax) as the impact of GW190814's rather precise $m_2$ constraint is likely to be more important.

We begin with the joint posterior for \Mmax~and GW190814's $m_2$,
\begin{widetext}
\begin{align}
    p(\Mmax, m_2|\data_{190814}, \{\data\}_{\mathcal{E} \setminus 190814}) & \propto p(\Mmax) p(m_2|\Mmax) p(\{\data\}_{\mathcal{E} \setminus 190814}|\Mmax) p(\data_{190814}|m_2) \nonumber \\
        & = p(\Mmax) p(\{\data\}_{\mathcal{E} \setminus 190814}|\Mmax) \left(\frac{p(m_2)\mathrm{H}(m_2\leq\Mmax)}{\int dm\, p(m)\mathrm{H}(m\leq\Mmax)}\right) p(\data_{190814}|m_2) \nonumber \\
        & \propto p(\Mmax|\{\data\}_{\mathcal{E} \setminus 190814}) \frac{\mathrm{H}(m_2\leq\Mmax)}{\int dm\, p(m) \mathrm{H}(m\leq\Mmax)} p(m_2|\data_{190814}) ,
\end{align}
where we have used the identity $p(m_2|\Mmax) = p(m_2) \mathrm{H}(m_2 \leq \Mmax) / \int dm\, p(m)\mathrm{H}(m\leq\Mmax)$.
If we wish to examine just the updated posterior on \Mmax, we can marginalize over $m_2$ to obtain
\begin{equation}
    p(\Mmax|\{\data\}_\mathcal{E}) \propto p(\Mmax|\{\data\}_{\mathcal{E} \setminus 190814}) \frac{\int dm_2\, p(m_2|\data_{190814}) \mathrm{H}(m_2\leq\Mmax)}{\int dm\, p(m) \mathrm{H}(m\leq\Mmax)}
\end{equation}
\end{widetext}
which is equivalent to the expression used to incorporate data from massive pulsars in \citet{Landry2020}.
The prior normalization acts as an Occam factor that favors \Mmax~only slightly larger than the observed $m_2$.
This term only enters if we assume \textit{a priori} that $m_2$ is a NS.
If we do not make this assumption, then the prior follows $d\mathcal{N}_\mathrm{tot}/dm_2$ insetad of $p(m_2|\Mmax)$.

Similarly, we can marginalize over \Mmax~to examine the resulting uncertainty in $m_2$.
\begin{widetext}
\begin{align}
    p(m_2| & \{\data\}_\mathcal{E}) \propto p(m_2|\data_{190814}) p(m_2) \int d\Mmax\, \frac{p(\Mmax|\{\data\}_{\mathcal{E} \setminus 190814})\mathrm{H}(m_2\leq\Mmax)}{\int dm\, p(m) \mathrm{H}(m\leq\Mmax)} .
\end{align}
\end{widetext}
We see that the Occam factor appears again to modify the distribution of \Mmax~from the $N-1$ other events.
This effectively modifies the prior $p(m_2)$ to only include values below $\Mmax$, subject to our uncertainty in $\Mmax$.
Recall that we have neglected selection effects in this section, meaning the fact that we have observed an $N^\mathrm{th}$ system is not relevant.

\begin{figure}
    \includegraphics[width=\columnwidth, clip=True, trim=0.1cm 0.3cm 0.1cm 0.1cm]{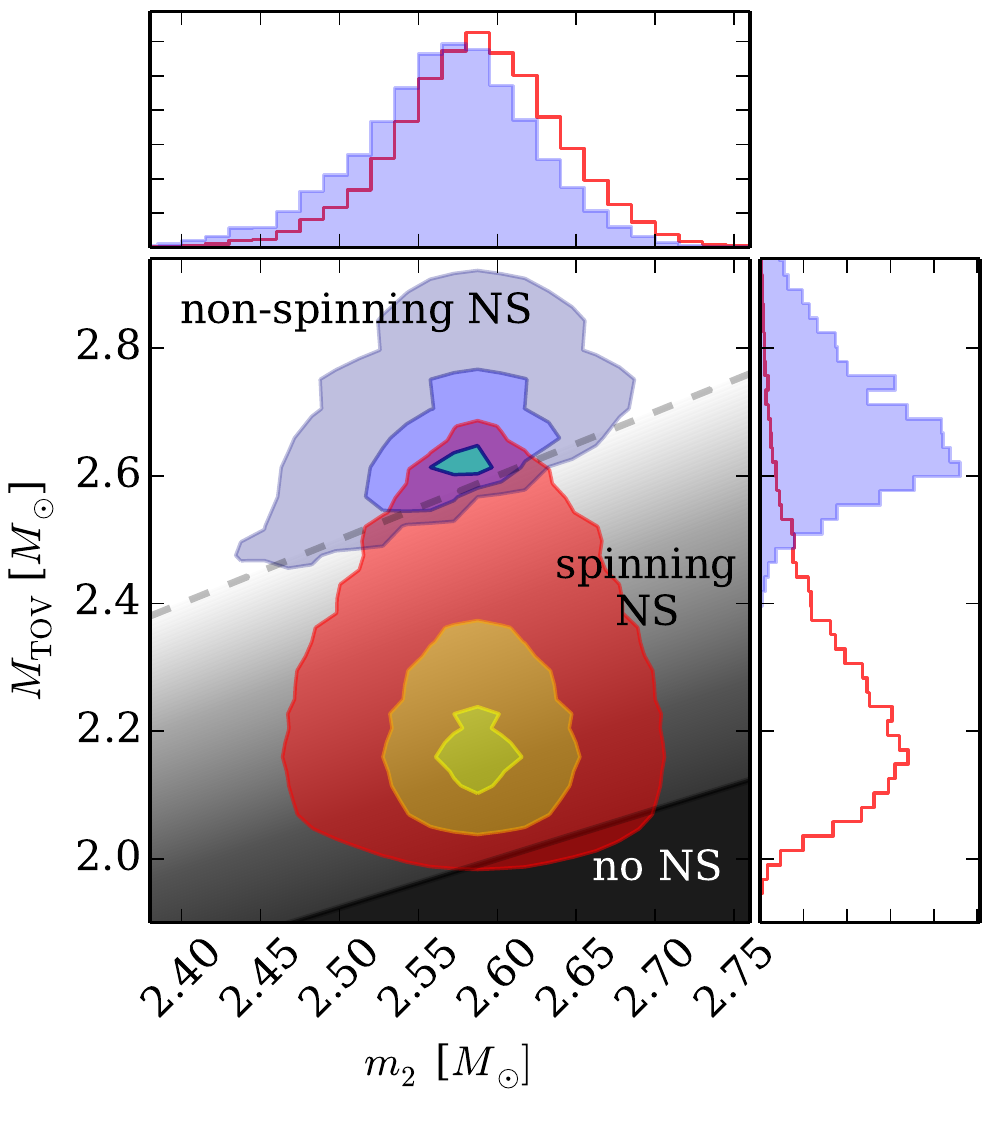}
    \caption{
        Distributions for GW190814's $m_2$ and \Mtov~\citep[LEC;][]{Landry2020} when making no assumptions about whether $m_2$ is a NS or a BH (\emph{red}) as well as the distributions when we assume \textit{a priori} that $m_2$ is a non-spinning NS (\emph{blue}).
        This analysis assumes a flat prior on $m_2$ that ends sharply at \Mtov.
        We note that our knowledge of $m_2$ is slightly shifted to lower values, but the main effect is to shift the \Mtov~posterior to larger values, although the width of the posterior remains nearly the same.
    }
    \label{fig:mmax-m2-NS}
\end{figure}

Fig.~\ref{fig:mmax-m2-NS} shows the results when we additionally assume GW190814's $m_2$ was slowly spinning so that $\Mmax=\Mtov$.
We assume flat priors on GW190814's component masses for simplicity, as the posterior only depends weakly on the population model (see Table~\ref{tab:GW190814}), subject to the constraint that $m_2 \leq \Mtov$.
Interestingly, we see that our knowledge of $m_2$ is not much improved, although it is shifted to slightly lower masses.
Instead, the main effect of the joint inference is to retain only the tail of the \Mmax~distribution.
As this tail is nearly a power law, the size of the 90\% highest-probability-density credible region only decreases by $\sim25\%$, but the median is shifted above the previous 90\% credible region's upper limit.
Note that $m_2$ and $\Mtov$ are no longer independent in the joint inference because we assume \textit{a priori} that $m_2 \leq \Mtov$.

Table~\ref{tab:updated m2 and Mmax} lists the original credible regions for both GW190814's $m_2$ and uncertainty in the NS \EOS~from \citet{Landry2020} as well as updated constraints obtained from this joint inference.
We note that our knowledge of the pressure at even relatively low densities is somewhat affected by this inference, as it is difficult to support non-spinning NSs as large as $m_2$ without an exceptionally stiff \EOS.
Indeed, the pressure at nuclear saturation density ($\rho_0$) is pushed to lower values in order match existing constraints on the radius and tidal deformability for stars with $m\sim1.4\,M_\odot$ from GW170817 and X-ray timing observations.
The value of the canonical radius is nearly unchanged, and the canonical tidal deformability changes by a smaller amount than the pressure at $2\rho_0$.

We do not present results obtained when assuming $\Mmax=1.3\Mtov$ as these are essentially identical to the original constraints.
To put that another way, GW190814's $m_2$ is barely consistent with a non-spinning NS, and therefore could impact our knowledge of the \EOS, but it is perfectly consistent with a (possibly rapidly) spinning NS, in which case we cannot learn about the \EOS~without direct measurements of matter effects in the waveform.

\begin{table}
    \caption{
        Medians and 90\% highest-probability-density credible regions for GW190814's $m_2$ and observables derived from our uncertainty in the NS \EOS.
        We only consider a flat mass distribution for $m_2$ up to \Mtov, as the shape of the mass distribution was found in Sec.~\ref{sec:examples} not to significantly affect our knowledge of $m_2$.
        We also present the change in the medians divided by the original size of the 90\% highest-probability-density credible region ($\Delta$/CR), emphasizing that the \EOS~observables associated with the highest densities are the most affected.
    }
    \label{tab:updated m2 and Mmax}
    \begin{center}
    \begin{tabular}{cccc}
    \hline
        & original & $m_2 \leq \Mtov$ & $\Delta/\text{CR}$ \\
    \hline
    \hline
        GW190814's $m_2\, [M_\odot]$ & \result{2.588^{+0.087}_{-0.086}} & \result{2.569^{+0.087}_{-0.095}} & \result{-10\%} \\
        $\Mtov\, [M_\odot]$ & \result{2.22^{+0.30}_{-0.20}} & \result{2.67^{+0.23}_{-0.16}} & \result{+90\%} \\
        $R_{1.4}\, [\mathrm{km}]$ & \result{12.32^{+1.09}_{-1.47}} & \result{12.46^{+0.96}_{-1.10}} & \result{+5\%} \\
        $\Lambda_{1.4}$ & \result{451^{+241}_{-279}} & \result{540^{+248}_{-181}} & \result{+17\%} \\
        $p(\rho_0)\, [10^{33}\, \mathrm{dyn}/\mathrm{cm}^2]$ & \result{4.3^{+3.8}_{-4.0}} & \result{3.0^{+4.0}_{-2.8}} & \result{-17\%} \\
        $p(2\rho_0)\, [10^{34}\, \mathrm{dyn}/\mathrm{cm}^2]$ & \result{3.8^{+2.7}_{-2.9}} & \result{5.4^{+4.5}_{-3.9}} & \result{+29\%} \\
        $p(4\rho_0)\, [10^{35}\, \mathrm{dyn}/\mathrm{cm}^2]$ & \result{3.4^{+1.8}_{-1.2}} & \result{6.1^{+2.4}_{-1.9}} & \result{+90\%} \\
        $p(6\rho_0)\, [10^{35}\, \mathrm{dyn}/\mathrm{cm}^2]$ & \result{8.6^{+5.3}_{-4.3}} & \result{14.3^{+6.1}_{-9.1}} & \result{+59\%} \\
    \hline
    \end{tabular}
    \end{center}
\end{table}

All the results in this section come with the substantial caveat that we do not know that GW190814 was a NSBH coalescence, and therefore we cannot assert that $m_2$ was a NS \textit{a priori} without the possibility of substantially biasing our inference of the \EOS.
No EM counterpart was observed, although this in itself is inconclusive \citep{Coughlin2019}.
Given the uncertainties in the shape of the overall mass distribution, the presence and depth of possible mass gaps, and even whether the lower edge of such mass gaps are related to \Mmax, updated constraints on the \EOS~obtained by assuming $m_2$ was a non-spinning NS should be met with healthy skepticism.
Indeed, the analyses in Section~\ref{sec:examples} suggest that $m_2$ is most consistent with either a spinning NS or a BH, and finer resolution is limited by the poor constraints on $m_2$'s spin.


\section{Discussion}
\label{sec:discussion}

In this paper, we investigated what knowledge of the total rate density of low-mass compact binary coalescences and uncertainty in the maximum NS mass, \Mmax, can tell us about the nature of individual compact objects.
While several matter signatures in the GW waveform can distinguish between types of low-mass stellar remnants, we expect that there will be a population of events for which these signatures are inconclusive and for which we will only be able to distinguish between types of objects based on their masses.
Specifically, we showed that we can place an upper bound on the posterior probability that any object is a NS, but that different assumptions about the unknown prior odds between NSs and BHs below \Mmax~could reduce our confidence that any particular object is a NS.
As such, we can only definitively rule out, rather than establish, the presence of a NS in a given coalescence without further knowledge of the subpopulation mass distributions or direct observations of tidal signatures.
Our hierarchical Bayesian approach generalizes previous mass-based classification schemes by not only including the full posterior distributions with population-informed priors but also accounting for our uncertainty in the mass that separates the NS and BH classes.

Applying this to two recent detections, we find that GW190425 was likely a BNS coalescence rather than a NSBH.
Because of the signal's relatively low S/N and correspondingly broad uncertainty in the component masses, we find that different assumptions about the astrophysical distribution of masses in this range can affect our confidence more than the systematic uncertainty between different \Mmax~distributions.
Most populations we assume favor symmetric mass ratios and therefore smaller values for $m_1$.
We typically find $P(m_1\leq\Mmax)$ between \result{70\%$ and $99\%}.
Although we cannot definitely prove that GW190425 did not involve a BH, and other plausible astrophysical scenarios have been proposed \citep{Foley2020, Han_2020}, this is nonetheless suggestive.

Similarly, we find GW190814's $m_2$ was almost certainly not a non-spinning NS, as \result{P(m_2 \leq \Mtov) \lesssim 6\%} for \Mtov~distributions based on NS \EOS~studies.
While we find that $m_2$ is completely consistent with a NS spinning near its break-up frequency, we also note that the data does not constrain $m_2$'s spin.
This agrees with the analysis presented in \citet{GW190814}.
Any higher-dimensional inference will be dominated by assumptions about the spin distribution.
For example, \citet{Most2020} place lower bounds on \Mtov~and the secondary's spin under the assumption that the object must lie near the universal relation for $\Mmax(\Mtov, \chi)$ \citep{Breu2016}, motivated by the belief that it could not have accreted much mass after it was initially formed.

These GW events emphasize the different limiting outcomes that can be expected in such an inference, when the knowledge of component masses is either significantly larger than or smaller than the uncertainty in \Mmax.
Unsurprisingly, the systematic uncertainty associated with whichever distribution is larger dominates the uncertainty in our conclusions.
As such, even perfect knowledge of the component masses for a particular event will not remove all systematic uncertainty, as several estimates for \Mmax~exist.
At the same time, perfect knowledge of \Mmax~simply means we will be limited by our understanding of how the population of low-mass stellar remnants is distributed.
This uncertainty could have a significant impact on our conclusions and should not be neglected.

Although universal relations that connect the maximum mass of non-spinning NSs, \Mtov, and how much more mass can be supported by the object's spin, it is not known how reliable these relations are in the presence of strong phase transitions.
Improving our theoretical understanding of this separatrix is unlikely to improve our understanding of GW190425, but could be useful for GW190814.
Specifically, using the universal relation for $\Mmax(\Mtov, \chi)$ from \citet{Breu2016}, we argue that a mass that is almost surely above \Mtov~may suggest the object could not have been a NS, as \Mmax~may only significantly exceed \Mtov~when the spins reach implausibly large amplitudes.
\citet{Most2020} arrives at a similar conclusion; assuming $m_2$ was a NS, they find it may have been the fasted spinning NS ever observed.

We also present a few caveats to keep in mind for this type of analysis.
Foremost is the fact that, although it may intuitively seem sensible to base classification on a Bayes factor to minimize the possible impact of prior beliefs, we show that Bayes factors are sensitive to the high-mass behavior of the BH mass distribution, which should be irrelevant for low-mass objects.
Using the posterior odds, or odds ratios, avoids this shortcoming as it only depends on the prior within the likelihood's support.
Furthermore, real binaries may come from a variety of formation channels, each of which may produce different distributions for BNSs, BBHs, and NSBH systems.
These formation channels may limit the maximum mass attained by astrophysical NSs to $m \leq \Mform \leq \Mmax$.
We argue that we are likely to only measure the sum of these distributions directly from the data and make the simplifying assumption that $\Mform=\Mmax$.
More detailed knowledge of individual subpopulations would likely be extremely useful.
\citet{Mandel2015} and \citet{10.1093/mnras/stw2883} investigate such knowledge from a single population synthesis calculation, finding that precise knowledge of subpopulations could obviate classification problem, allowing most systems to be identified ``by eye'' or with simple clustering algorithms from the masses alone without the need for the formal machinery developed here.

Looking forward, \citet{MatterMatters} suggests that we should expect as many as one out of every six detections to involve a primary mass $\lesssim7\,M_\odot$, and that a significant fraction of these may have ambiguous classifications based on their masses alone.
Indeed, existing public alerts from O3 \citep{gracedb} include several candidates classified as likely to contain possibly ambiguous components \citep[BNS, NSBH, and Mass Gap events as defined in][]{userguide}.
We therefore expect the statistical and systematic uncertainties explored here to remain relevant throughout the advanced detector era, although \citet{ChenJohnsonMcDaniel2020} show that we should expect to relatively confidently detect tidal signatures for nearby systems with the expected O4 detector sensitivities.

Indeed, although the S/N distribution of compact binary systems observed with third-generation detectors will peak above the detection threshold \citep[see Fig. 7 of][]{Vitale2016} and a much larger fraction of detections will have clearly discernible matter signatures within their waveforms, a nontrivial fraction will still have low enough S/N that their masses may be our best way to identify their constituents.

Nonetheless, even bearing in mind the systematic uncertainties from our imperfect knowledge of the distribution of low-mass stellar remnants and the \EOS~of dense nuclear matter, it is remarkable that GW observations already allow us to ask such pointed questions about individual astrophysical objects so soon after the first direct detection of GWs \citep{GW150914}.
This demonstrates the vast amount of information encoded within GW signals and the unprecedented opportunities they provide to learn about astrophysical population of compact objects.


\acknowledgements

The authors thank Jolien Creighton, Maya Fishbach, and Daniel Holz for their feedback while preparing this manuscript.
The authors are also grateful for useful discussions with Katerina Chatziioannou and the broader LIGO-Virgo-KAGRA Extreme Matter and Rates \& Populations working groups.
R.~E. is supported at the University of Chicago by the Kavli Institute for Cosmological Physics through an endowment from the Kavli Foundation and its founder Fred Kavli.
P.~L. is supported by National Science Foundation award PHY-1836734 and by a gift from the Black Family Trust to the Gravitational-Wave Physics \& Astronomy Center.
The authors also gratefully acknowledge the computational resources provided by the LIGO Laboratory and supported by NSF grants PHY-0757058 and PHY-0823459.


\bibliography{refs}

\begin{thebibliography}{}
\expandafter\ifx\csname natexlab\endcsname\relax\def\natexlab#1{#1}\fi
\providecommand{\url}[1]{\href{#1}{#1}}
\providecommand{\dodoi}[1]{doi:~\href{http://doi.org/#1}{\nolinkurl{#1}}}
\providecommand{\doeprint}[1]{\href{http://ascl.net/#1}{\nolinkurl{http://ascl.net/#1}}}
\providecommand{\doarXiv}[1]{\href{https://arxiv.org/abs/#1}{\nolinkurl{https://arxiv.org/abs/#1}}}

\bibitem[{Aasi {et~al.}(2015)}]{AdvancedVirgo}
Aasi, J., {et~al.} 2015, Class. Quant. Grav., 32, 074001,
  \dodoi{10.1088/0264-9381/32/7/074001}

\bibitem[{Abbott {et~al.}(2016)}]{GW150914}
Abbott, B.~P., {et~al.} 2016, Phys. Rev. Lett., 116, 061102,
  \dodoi{10.1103/PhysRevLett.116.061102}

\bibitem[{Abbott {et~al.}(2017{\natexlab{a}})}]{GW170817}
---. 2017{\natexlab{a}}, Phys. Rev. Lett., 119, 161101,
  \dodoi{10.1103/PhysRevLett.119.161101}

\bibitem[{Abbott {et~al.}(2017{\natexlab{b}})}]{GW170817-MMA}
---. 2017{\natexlab{b}}, The Astrophysical Journal, 848, L12,
  \dodoi{10.3847/2041-8213/aa91c9}

\bibitem[{Abbott {et~al.}(2017{\natexlab{c}})}]{GW170817-KN}
---. 2017{\natexlab{c}}, The Astrophysical Journal, 850, L39,
  \dodoi{10.3847/2041-8213/aa9478}

\bibitem[{Abbott {et~al.}(2018)}]{GW170817-EOS}
---. 2018, Phys. Rev. Lett., 121, 161101,
  \dodoi{10.1103/PhysRevLett.121.161101}

\bibitem[{{Abbott} {et~al.}(2019){Abbott}, {Abbott}, {Abbott}, {Acernese},
  {Ackley}, {Adams}, {Adams}, {Addesso}, {Adhikari}, {Adya}, \&
  et~al.}]{GW170817-Props}
{Abbott}, B.~P., {Abbott}, R., {Abbott}, T.~D., {et~al.} 2019, Physical Review
  X, 9, 011001, \dodoi{10.1103/PhysRevX.9.011001}

\bibitem[{Abbott {et~al.}(2019{\natexlab{a}})}]{GW170817NonlinearTides}
Abbott, B.~P., {et~al.} 2019{\natexlab{a}}, Phys. Rev. Lett., 122, 061104,
  \dodoi{10.1103/PhysRevLett.122.061104}

\bibitem[{Abbott {et~al.}(2019{\natexlab{b}})}]{GWTC-1}
---. 2019{\natexlab{b}}, Phys. Rev. X, 9, 031040,
  \dodoi{10.1103/PhysRevX.9.031040}

\bibitem[{Abbott {et~al.}(2019{\natexlab{c}})}]{GWTC1RatesAndPopulations}
---. 2019{\natexlab{c}}, The Astrophysical Journal, 882, L24,
  \dodoi{10.3847/2041-8213/ab3800}

\bibitem[{{Abbott} {et~al.}(2020){Abbott}, {Abbott}, {Abbott}, {Abraham},
  {Acernese}, {Ackley}, {Adams}, {Adhikari}, {Adya}, {Affeldt}, \&
  et~al.}]{GW190425}
{Abbott}, B.~P., {Abbott}, R., {Abbott}, T.~D., {et~al.} 2020, \apjl, 892, L3,
  \dodoi{10.3847/2041-8213/ab75f5}

\bibitem[{Abbott {et~al.}(2020{\natexlab{a}})}]{GW170817-ModelSelection}
Abbott, B.~P., {et~al.} 2020{\natexlab{a}}, Classical and Quantum Gravity, 37,
  045006, \dodoi{10.1088/1361-6382/ab5f7c}

\bibitem[{Abbott {et~al.}(2020{\natexlab{b}})}]{GW190814}
Abbott, R., {et~al.} 2020{\natexlab{b}}, The Astrophysical Journal, 896, L44,
  \dodoi{10.3847/2041-8213/ab960f}

\bibitem[{Acernese {et~al.}(2015)}]{AdvancedLIGO}
Acernese, F., {et~al.} 2015, Class. Quant. Grav., 32, 024001,
  \dodoi{10.1088/0264-9381/32/2/024001}

\bibitem[{Ai {et~al.}(2020)Ai, Gao, \& Zhang}]{Ai_2020}
Ai, S., Gao, H., \& Zhang, B. 2020, The Astrophysical Journal, 893, 146,
  \dodoi{10.3847/1538-4357/ab80bd}

\bibitem[{Alsing {et~al.}(2018)Alsing, Silva, \& Berti}]{Alsing2018}
Alsing, J., Silva, H.~O., \& Berti, E. 2018, Monthly Notices of the Royal
  Astronomical Society, 478, 1377, \dodoi{10.1093/mnras/sty1065}

\bibitem[{Antoniadis {et~al.}(2013)Antoniadis, Freire, Wex, Tauris, Lynch,
  {et~al.}}]{Antoniadis:2013pzd}
Antoniadis, J., Freire, P.~C., Wex, N., {et~al.} 2013, Science, 340, 1233232,
  \dodoi{10.1126/science.1233232}

\bibitem[{{Antoniadis} {et~al.}(2016){Antoniadis}, {Tauris}, {Ozel}, {Barr},
  {Champion}, \& {Freire}}]{AntoniadisTauris2016}
{Antoniadis}, J., {Tauris}, T.~M., {Ozel}, F., {et~al.} 2016.
\newblock \doarXiv{1605.01665}

\bibitem[{Bailyn {et~al.}(1998)Bailyn, Jain, Coppi, \& Orosz}]{Bailyn_1998}
Bailyn, C.~D., Jain, R.~K., Coppi, P., \& Orosz, J.~A. 1998, The Astrophysical
  Journal, 499, 367, \dodoi{10.1086/305614}

\bibitem[{{Barbieri} {et~al.}(2020){Barbieri}, {Salafia}, {Colpi}, {Ghirlanda},
  \& {Perego}}]{BarbieriSalafia2020}
{Barbieri}, C., {Salafia}, O.~S., {Colpi}, M., {Ghirlanda}, G., \& {Perego}, A.
  2020, arXiv e-prints, arXiv:2002.09395.
\newblock \doarXiv{2002.09395}

\bibitem[{{Barbieri} {et~al.}(2019){Barbieri}, {Salafia}, {Perego}, {Colpi}, \&
  {Ghirlanda}}]{BarbieriSalafia2019}
{Barbieri}, C., {Salafia}, O.~S., {Perego}, A., {Colpi}, M., \& {Ghirlanda}, G.
  2019, \aap, 625, A152, \dodoi{10.1051/0004-6361/201935443}

\bibitem[{Bauswein {et~al.}(2013)Bauswein, Baumgarte, \&
  Janka}]{PhysRevLett.111.131101}
Bauswein, A., Baumgarte, T.~W., \& Janka, H.-T. 2013, Phys. Rev. Lett., 111,
  131101, \dodoi{10.1103/PhysRevLett.111.131101}

\bibitem[{Belczynski {et~al.}(2012)Belczynski, Wiktorowicz, Fryer, Holz, \&
  Kalogera}]{Belczynski_2012}
Belczynski, K., Wiktorowicz, G., Fryer, C.~L., Holz, D.~E., \& Kalogera, V.
  2012, The Astrophysical Journal, 757, 91, \dodoi{10.1088/0004-637x/757/1/91}

\bibitem[{Breu \& Rezzolla(2016)}]{Breu2016}
Breu, C., \& Rezzolla, L. 2016, Monthly Notices of the Royal Astronomical
  Society, 459, 646, \dodoi{10.1093/mnras/stw575}

\bibitem[{Bulik {et~al.}(2003)Bulik, Belczynski, \&
  Kalogera}]{10.1117/12.458968}
Bulik, T., Belczynski, K., \& Kalogera, V. 2003, in Gravitational-Wave
  Detection, ed. P.~Saulson \& A.~M. Cruise, Vol. 4856, International Society
  for Optics and Photonics (SPIE), 146 -- 155, \dodoi{10.1117/12.458968}

\bibitem[{{Cardoso} \& {Pani}(2019)}]{CardosoPani2019}
{Cardoso}, V., \& {Pani}, P. 2019, Living Reviews in Relativity, 22, 4,
  \dodoi{10.1007/s41114-019-0020-4}

\bibitem[{Carney {et~al.}(2018)Carney, Wade, \& Irwin}]{Carney2018}
Carney, M.~F., Wade, L.~E., \& Irwin, B.~S. 2018, Phys. Rev. D, 98, 063004,
  \dodoi{10.1103/PhysRevD.98.063004}

\bibitem[{{Carr} \& {Hawking}(1974)}]{CarrHawking1974}
{Carr}, B.~J., \& {Hawking}, S.~W. 1974, \mnras, 168, 399,
  \dodoi{10.1093/mnras/168.2.399}

\bibitem[{{Chatziioannou} \& {Farr}(2020)}]{Chatziioannou2020}
{Chatziioannou}, K., \& {Farr}, W.~M. 2020, arXiv e-prints, arXiv:2005.00482.
\newblock \doarXiv{2005.00482}

\bibitem[{{Chen} {et~al.}(2020){Chen}, {Johnson-McDaniel}, {Dietrich}, \&
  {Dudi}}]{ChenJohnsonMcDaniel2020}
{Chen}, A., {Johnson-McDaniel}, N.~K., {Dietrich}, T., \& {Dudi}, R. 2020,
  \prd, 101, 103008, \dodoi{10.1103/PhysRevD.101.103008}

\bibitem[{Chen \& Chatziioannou(2020)}]{Chen2020}
Chen, H.-Y., \& Chatziioannou, K. 2020, The Astrophysical Journal, 893, L41,
  \dodoi{10.3847/2041-8213/ab86bc}

\bibitem[{{Chen} \& {Chatziioannou}(2020)}]{ChenChatziioannou2020}
{Chen}, H.-Y., \& {Chatziioannou}, K. 2020, \apjl, 893, L41,
  \dodoi{10.3847/2041-8213/ab86bc}

\bibitem[{{Cook} {et~al.}(1994){Cook}, {Shapiro}, \&
  {Teukolsky}}]{1994ApJ...424..823C}
{Cook}, G.~B., {Shapiro}, S.~L., \& {Teukolsky}, S.~A. 1994, \apj, 424, 823,
  \dodoi{10.1086/173934}

\bibitem[{{Coughlin} \& {Dietrich}(2019)}]{CoughlinDietrich2019}
{Coughlin}, M.~W., \& {Dietrich}, T. 2019, \prd, 100, 043011,
  \dodoi{10.1103/PhysRevD.100.043011}

\bibitem[{Coughlin {et~al.}(2019)Coughlin, Dietrich, Antier, Bulla, Foucart,
  Hotokezaka, Raaijmakers, Hinderer, \& Nissanke}]{Coughlin2019}
Coughlin, M.~W., Dietrich, T., Antier, S., {et~al.} 2019, Monthly Notices of
  the Royal Astronomical Society, 492, 863, \dodoi{10.1093/mnras/stz3457}

\bibitem[{{Cromartie} {et~al.}(2020){Cromartie}, {Fonseca}, {Ransom},
  {Demorest}, {Arzoumanian}, {et~al.}}]{Cromartie:2019kug}
{Cromartie}, H.~T., {Fonseca}, E., {Ransom}, S.~M., {et~al.} 2020, Nature
  Astronomy, 4, 72, \dodoi{10.1038/s41550-019-0880-2}

\bibitem[{{Datta} {et~al.}(2020){Datta}, {Phukon}, \& {Bose}}]{DattaPhukon2020}
{Datta}, S., {Phukon}, K.~S., \& {Bose}, S. 2020, arXiv e-prints,
  arXiv:2004.05974.
\newblock \doarXiv{2004.05974}

\bibitem[{Dietrich {et~al.}(2020)Dietrich, Coughlin, Pang, Bulla, Heinzel,
  Issa, Tews, \& Antier}]{Dietrich2020}
Dietrich, T., Coughlin, M.~W., Pang, P. T.~H., {et~al.} 2020.
\newblock \doarXiv{2002.11355}

\bibitem[{Ertl {et~al.}(2020)Ertl, Woosley, Sukhbold, \& Janka}]{Ertl_2020}
Ertl, T., Woosley, S.~E., Sukhbold, T., \& Janka, H.-T. 2020, The Astrophysical
  Journal, 890, 51, \dodoi{10.3847/1538-4357/ab6458}

\bibitem[{Essick {et~al.}(2020)Essick, Landry, \& Holz}]{Essick2020}
Essick, R., Landry, P., \& Holz, D.~E. 2020, Phys. Rev. D, 101, 063007,
  \dodoi{10.1103/PhysRevD.101.063007}

\bibitem[{Essick {et~al.}(2016)Essick, Vitale, \&
  Weinberg}]{PhysRevD.94.103012}
Essick, R., Vitale, S., \& Weinberg, N.~N. 2016, Phys. Rev. D, 94, 103012,
  \dodoi{10.1103/PhysRevD.94.103012}

\bibitem[{{Farr} \& {Chatziioannou}(2020)}]{FarrChatziioannou2020}
{Farr}, W.~M., \& {Chatziioannou}, K. 2020, Research Notes of the American
  Astronomical Society, 4, 65, \dodoi{10.3847/2515-5172/ab9088}

\bibitem[{Farr {et~al.}(2015)Farr, Gair, Mandel, \&
  Cutler}]{PhysRevD.91.023005}
Farr, W.~M., Gair, J.~R., Mandel, I., \& Cutler, C. 2015, Phys. Rev. D, 91,
  023005, \dodoi{10.1103/PhysRevD.91.023005}

\bibitem[{Farr {et~al.}(2011)Farr, Sravan, Cantrell, Kreidberg, Bailyn, Mandel,
  \& Kalogera}]{Farr_2011}
Farr, W.~M., Sravan, N., Cantrell, A., {et~al.} 2011, The Astrophysical
  Journal, 741, 103, \dodoi{10.1088/0004-637x/741/2/103}

\bibitem[{Farrow {et~al.}(2019)Farrow, Zhu, \& Thrane}]{Farrow2019}
Farrow, N., Zhu, X.-J., \& Thrane, E. 2019, The Astrophysical Journal, 876, 18,
  \dodoi{10.3847/1538-4357/ab12e3}

\bibitem[{{Fasano} {et~al.}(2020){Fasano}, {Wong}, {Maselli}, {Berti},
  {Ferrari}, \& {Sathyaprakash}}]{Fasano2020}
{Fasano}, M., {Wong}, K. W.~K., {Maselli}, A., {et~al.} 2020, arXiv e-prints,
  arXiv:2005.01726.
\newblock \doarXiv{2005.01726}

\bibitem[{{Fern{\'a}ndez} {et~al.}(2017){Fern{\'a}ndez}, {Foucart}, {Kasen},
  {Lippuner}, {Desai}, \& {Roberts}}]{FernandezFoucart2017}
{Fern{\'a}ndez}, R., {Foucart}, F., {Kasen}, D., {et~al.} 2017, Classical and
  Quantum Gravity, 34, 154001, \dodoi{10.1088/1361-6382/aa7a77}

\bibitem[{Fishbach {et~al.}(2020{\natexlab{a}})Fishbach, Essick, \&
  Holz}]{MatterMatters}
Fishbach, M., Essick, R., \& Holz, D.~E. 2020{\natexlab{a}}, arXiv e-prints,
  arXiv:2006.13178

\bibitem[{Fishbach {et~al.}(2020{\natexlab{b}})Fishbach, Farr, \&
  Holz}]{FishbachFarrHolz}
Fishbach, M., Farr, W.~M., \& Holz, D.~E. 2020{\natexlab{b}}, The Astrophysical
  Journal, 891, L31, \dodoi{10.3847/2041-8213/ab77c9}

\bibitem[{Fishbach \& Holz(2017)}]{Fishbach_2017}
Fishbach, M., \& Holz, D.~E. 2017, The Astrophysical Journal, 851, L25,
  \dodoi{10.3847/2041-8213/aa9bf6}

\bibitem[{Fishbach \& Holz(2020)}]{PickyPartners}
---. 2020, The Astrophysical Journal, 891, L27,
  \dodoi{10.3847/2041-8213/ab7247}

\bibitem[{Flanagan \& Hinderer(2008)}]{PhysRevD.77.021502}
Flanagan, E.~E., \& Hinderer, T. 2008, Phys. Rev. D, 77, 021502,
  \dodoi{10.1103/PhysRevD.77.021502}

\bibitem[{Foley {et~al.}(2020)Foley, Coulter, Kilpatrick, Piro, Ramirez-Ruiz,
  \& Schwab}]{Foley2020}
Foley, R.~J., Coulter, D.~A., Kilpatrick, C.~D., {et~al.} 2020, Monthly Notices
  of the Royal Astronomical Society, 494, 190, \dodoi{10.1093/mnras/staa725}

\bibitem[{{Foucart} {et~al.}(2018){Foucart}, {Hinderer}, \&
  {Nissanke}}]{FoucartHinderer2018}
{Foucart}, F., {Hinderer}, T., \& {Nissanke}, S. 2018, \prd, 98, 081501,
  \dodoi{10.1103/PhysRevD.98.081501}

\bibitem[{Fryer \& Kalogera(2001)}]{Fryer_2001}
Fryer, C.~L., \& Kalogera, V. 2001, The Astrophysical Journal, 554, 548,
  \dodoi{10.1086/321359}

\bibitem[{{Gopalan} {et~al.}(2015){Gopalan}, {Vrtilek}, \&
  {Bornn}}]{GopalanVrtilek2015}
{Gopalan}, G., {Vrtilek}, S.~D., \& {Bornn}, L. 2015, \apj, 809, 40,
  \dodoi{10.1088/0004-637X/809/1/40}

\bibitem[{{Gupta} {et~al.}(2020){Gupta}, {Gerosa}, {Arun}, {Berti}, {Farr}, \&
  {Sathyaprakash}}]{GuptaGerosa2020}
{Gupta}, A., {Gerosa}, D., {Arun}, K.~G., {et~al.} 2020, \prd, 101, 103036,
  \dodoi{10.1103/PhysRevD.101.103036}

\bibitem[{{Haensel} {et~al.}(1995){Haensel}, {Salgado}, \&
  {Bonazzola}}]{1995A&A...296..745H}
{Haensel}, P., {Salgado}, M., \& {Bonazzola}, S. 1995, \aap, 296, 745

\bibitem[{{Han} {et~al.}(2020){Han}, {Tang}, {Hu}, {Li}, {Jiang}, {Jin}, {Fan},
  \& {Wei}}]{HanTang2020}
{Han}, M.-Z., {Tang}, S.-P., {Hu}, Y.-M., {et~al.} 2020, \apjl, 891, L5,
  \dodoi{10.3847/2041-8213/ab745a}

\bibitem[{Han {et~al.}(2020)Han, Tang, Hu, Li, Jiang, Jin, Fan, \&
  Wei}]{Han_2020}
Han, M.-Z., Tang, S.-P., Hu, Y.-M., {et~al.} 2020, The Astrophysical Journal,
  891, L5, \dodoi{10.3847/2041-8213/ab745a}

\bibitem[{{Hannam} {et~al.}(2013){Hannam}, {Brown}, {Fairhurst}, {Fryer}, \&
  {Harry}}]{HannamBrown2013}
{Hannam}, M., {Brown}, D.~A., {Fairhurst}, S., {Fryer}, C.~L., \& {Harry},
  I.~W. 2013, \apjl, 766, L14, \dodoi{10.1088/2041-8205/766/1/L14}

\bibitem[{{Haster} {et~al.}(2020){Haster}, {Chatziioannou}, {Bauswein}, \&
  {Clark}}]{Haster2020}
{Haster}, C.-J., {Chatziioannou}, K., {Bauswein}, A., \& {Clark}, J.~A. 2020,
  arXiv e-prints, arXiv:2004.11334.
\newblock \doarXiv{2004.11334}

\bibitem[{{Hessels} {et~al.}(2006){Hessels}, {Ransom}, {Stairs}, {Freire},
  {Kaspi}, \& {Camilo}}]{Hessels2006}
{Hessels}, J. W.~T., {Ransom}, S.~M., {Stairs}, I.~H., {et~al.} 2006, Science,
  311, 1901, \dodoi{10.1126/science.1123430}

\bibitem[{Hinderer {et~al.}(2016)Hinderer, Taracchini, Foucart, Buonanno,
  Steinhoff, Duez, Kidder, Pfeiffer, Scheel, Szilagyi, Hotokezaka, Kyutoku,
  Shibata, \& Carpenter}]{PhysRevLett.116.181101}
Hinderer, T., Taracchini, A., Foucart, F., {et~al.} 2016, Phys. Rev. Lett.,
  116, 181101, \dodoi{10.1103/PhysRevLett.116.181101}

\bibitem[{{Hinderer} {et~al.}(2019){Hinderer}, {Nissanke}, {Foucart},
  {Hotokezaka}, {Vincent}, {Kasliwal}, {Schmidt}, {Williamson}, {Nichols},
  {Duez}, \& et~al.}]{HindererNissanke2019}
{Hinderer}, T., {Nissanke}, S., {Foucart}, F., {et~al.} 2019, \prd, 100,
  063021, \dodoi{10.1103/PhysRevD.100.063021}

\bibitem[{{Johnson-McDaniel} {et~al.}(2018){Johnson-McDaniel}, {Mukherjee},
  {Kashyap}, {Ajith}, {Del Pozzo}, \& {Vitale}}]{JohnsonMcDanielMukherjee2018}
{Johnson-McDaniel}, N.~K., {Mukherjee}, A., {Kashyap}, R., {et~al.} 2018, arXiv
  e-prints, arXiv:1804.08026.
\newblock \doarXiv{1804.08026}

\bibitem[{Kapadia {et~al.}(2020)Kapadia, Caudill, Creighton, Farr, Mendell,
  Weinstein, Cannon, Fong, Godwin, Lo, Magee, Meacher, Messick, Mohite,
  Mukherjee, \& Sachdev}]{Kapadia_2020}
Kapadia, S.~J., Caudill, S., Creighton, J. D.~E., {et~al.} 2020, Classical and
  Quantum Gravity, 37, 045007, \dodoi{10.1088/1361-6382/ab5f2d}

\bibitem[{{Krishnendu} {et~al.}(2017){Krishnendu}, {Arun}, \&
  {Mishra}}]{KrishnenduArun2017}
{Krishnendu}, N.~V., {Arun}, K.~G., \& {Mishra}, C.~K. 2017, \prl, 119, 091101,
  \dodoi{10.1103/PhysRevLett.119.091101}

\bibitem[{{Krishnendu} {et~al.}(2019){Krishnendu}, {Saleem}, {Samajdar},
  {Arun}, {Del Pozzo}, \& {Mishra}}]{KrishnenduSaleem2019}
{Krishnendu}, N.~V., {Saleem}, M., {Samajdar}, A., {et~al.} 2019, \prd, 100,
  104019, \dodoi{10.1103/PhysRevD.100.104019}

\bibitem[{{Kyutoku} {et~al.}(2020){Kyutoku}, {Fujibayashi}, {Hayashi},
  {Kawaguchi}, {Kiuchi}, {Shibata}, \& {Tanaka}}]{KyutokuFujibayashi2020}
{Kyutoku}, K., {Fujibayashi}, S., {Hayashi}, K., {et~al.} 2020, \apjl, 890, L4,
  \dodoi{10.3847/2041-8213/ab6e70}

\bibitem[{Lackey \& Wade(2015)}]{Lackey2015}
Lackey, B.~D., \& Wade, L. 2015, Phys. Rev. D, 91, 043002,
  \dodoi{10.1103/PhysRevD.91.043002}

\bibitem[{{Lai}(1994)}]{1994MNRAS.270..611L}
{Lai}, D. 1994, \mnras, 270, 611, \dodoi{10.1093/mnras/270.3.611}

\bibitem[{{Landry} {et~al.}(2020){Landry}, {Essick}, \&
  {Chatziioannou}}]{Landry2020}
{Landry}, P., {Essick}, R., \& {Chatziioannou}, K. 2020, arXiv e-prints,
  arXiv:2003.04880.
\newblock \doarXiv{2003.04880}

\bibitem[{{Lattimer} \& {Prakash}(2001)}]{Lattimer2001}
{Lattimer}, J.~M., \& {Prakash}, M. 2001, \apj, 550, 426,
  \dodoi{10.1086/319702}

\bibitem[{{Littenberg} {et~al.}(2015){Littenberg}, {Farr}, {Coughlin},
  {Kalogera}, \& {Holz}}]{Littenberg2015}
{Littenberg}, T.~B., {Farr}, B., {Coughlin}, S., {Kalogera}, V., \& {Holz},
  D.~E. 2015, \apjl, 807, L24, \dodoi{10.1088/2041-8205/807/2/L24}

\bibitem[{{Loredo}(2004)}]{2004AIPC..735..195L}
{Loredo}, T.~J. 2004, in American Institute of Physics Conference Series, Vol.
  735, American Institute of Physics Conference Series, ed. R.~{Fischer},
  R.~{Preuss}, \& U.~V. {Toussaint}, 195--206, \dodoi{10.1063/1.1835214}

\bibitem[{{Loredo} \& {Wasserman}(1995)}]{1995ApJS...96..261L}
{Loredo}, T.~J., \& {Wasserman}, I.~M. 1995, \apjs, 96, 261,
  \dodoi{10.1086/192119}

\bibitem[{{Mandel} \& {Farmer}(2018)}]{2018arXiv180605820M}
{Mandel}, I., \& {Farmer}, A. 2018, arXiv e-prints, arXiv:1806.05820.
\newblock \doarXiv{1806.05820}

\bibitem[{Mandel {et~al.}(2016)Mandel, Farr, Colonna, Stevenson, Tiňo, \&
  Veitch}]{10.1093/mnras/stw2883}
Mandel, I., Farr, W.~M., Colonna, A., {et~al.} 2016, Monthly Notices of the
  Royal Astronomical Society, 465, 3254, \dodoi{10.1093/mnras/stw2883}

\bibitem[{Mandel {et~al.}(2019)Mandel, Farr, \& Gair}]{10.1093/mnras/stz896}
Mandel, I., Farr, W.~M., \& Gair, J.~R. 2019, Monthly Notices of the Royal
  Astronomical Society, 486, 1086, \dodoi{10.1093/mnras/stz896}

\bibitem[{{Mandel} {et~al.}(2015){Mandel}, {Haster}, {Dominik}, \&
  {Belczynski}}]{Mandel2015}
{Mandel}, I., {Haster}, C.-J., {Dominik}, M., \& {Belczynski}, K. 2015, \mnras,
  450, L85, \dodoi{10.1093/mnrasl/slv054}

\bibitem[{Margalit \& Metzger(2017)}]{Margalit_2017}
Margalit, B., \& Metzger, B.~D. 2017, The Astrophysical Journal, 850, L19,
  \dodoi{10.3847/2041-8213/aa991c}

\bibitem[{Margalit \& Metzger(2019)}]{Margalit_2019}
---. 2019, The Astrophysical Journal, 880, L15,
  \dodoi{10.3847/2041-8213/ab2ae2}

\bibitem[{{Meszaros}(1974)}]{Meszaros1974}
{Meszaros}, P. 1974, \aap, 37, 225

\bibitem[{Miller {et~al.}(2019{\natexlab{a}})Miller, Chirenti, \&
  Lamb}]{Miller2019b}
Miller, M.~C., Chirenti, C., \& Lamb, F.~K. 2019{\natexlab{a}}, The
  Astrophysical Journal, 888, 12, \dodoi{10.3847/1538-4357/ab4ef9}

\bibitem[{Miller {et~al.}(2019{\natexlab{b}})}]{Miller2019}
Miller, M.~C., {et~al.} 2019{\natexlab{b}}, Astrophys. J. Lett., 887, L24.
\newblock \doarXiv{1912.05705}

\bibitem[{Most {et~al.}(2020)Most, Papenfort, Weih, \& Rezzolla}]{Most2020}
Most, E.~R., Papenfort, L.~J., Weih, L.~R., \& Rezzolla, L. 2020, arXiv
  e-prints, arXiv:2006.14601

\bibitem[{{{\"O}zel} {et~al.}(2010){{\"O}zel}, {Psaltis}, {Narayan}, \&
  {McClintock}}]{2010ApJ...725.1918O}
{{\"O}zel}, F., {Psaltis}, D., {Narayan}, R., \& {McClintock}, J.~E. 2010,
  \apj, 725, 1918, \dodoi{10.1088/0004-637X/725/2/1918}

\bibitem[{Postnov \& Yungelson(2014)}]{Postnov2014}
Postnov, K.~A., \& Yungelson, L.~R. 2014, Living Reviews in Relativity, 1433,
  \dodoi{10.12942/lrr-2014-3}

\bibitem[{Raaijmakers {et~al.}(2019)}]{Raaijmakers2019}
Raaijmakers, G., {et~al.} 2019, Astrophys. J. Lett., 887, L22,
  \dodoi{10.3847/2041-8213/ab451a}

\bibitem[{Read {et~al.}(2009)Read, Lackey, Owen, \&
  Friedman}]{PhysRevD.79.124032}
Read, J.~S., Lackey, B.~D., Owen, B.~J., \& Friedman, J.~L. 2009, Phys. Rev. D,
  79, 124032, \dodoi{10.1103/PhysRevD.79.124032}

\bibitem[{{Reisenegger} \& {Goldreich}(1994)}]{1994ApJ...426..688R}
{Reisenegger}, A., \& {Goldreich}, P. 1994, \apj, 426, 688,
  \dodoi{10.1086/174105}

\bibitem[{Rezzolla {et~al.}(2018)Rezzolla, Most, \& Weih}]{Rezzolla_2018}
Rezzolla, L., Most, E.~R., \& Weih, L.~R. 2018, The Astrophysical Journal, 852,
  L25, \dodoi{10.3847/2041-8213/aaa401}

\bibitem[{{Rhoades} \& {Ruffini}(1974)}]{RhoadesRuffini1974}
{Rhoades}, C.~E., \& {Ruffini}, R. 1974, \prl, 32, 324,
  \dodoi{10.1103/PhysRevLett.32.324}

\bibitem[{Riley {et~al.}(2019)}]{Riley2019}
Riley, T.~E., {et~al.} 2019, Astrophys. J. Lett., 887, L21,
  \dodoi{10.3847/2041-8213/ab481c}

\bibitem[{Rodriguez {et~al.}(2018)Rodriguez, Amaro-Seoane, Chatterjee, Kremer,
  Rasio, Samsing, Ye, \& Zevin}]{PhysRevD.98.123005}
Rodriguez, C.~L., Amaro-Seoane, P., Chatterjee, S., {et~al.} 2018, Phys. Rev.
  D, 98, 123005, \dodoi{10.1103/PhysRevD.98.123005}

\bibitem[{Rodriguez {et~al.}(2019)Rodriguez, Zevin, Amaro-Seoane, Chatterjee,
  Kremer, Rasio, \& Ye}]{PhysRevD.100.043027}
Rodriguez, C.~L., Zevin, M., Amaro-Seoane, P., {et~al.} 2019, Phys. Rev. D,
  100, 043027, \dodoi{10.1103/PhysRevD.100.043027}

\bibitem[{{Shibata} \& {Taniguchi}(2008)}]{ShibataTaniguchi2008}
{Shibata}, M., \& {Taniguchi}, K. 2008, \prd, 77, 084015,
  \dodoi{10.1103/PhysRevD.77.084015}

\bibitem[{Shibata {et~al.}(2019)Shibata, Zhou, Kiuchi, \&
  Fujibayashi}]{PhysRevD.100.023015}
Shibata, M., Zhou, E., Kiuchi, K., \& Fujibayashi, S. 2019, Phys. Rev. D, 100,
  023015, \dodoi{10.1103/PhysRevD.100.023015}

\bibitem[{Steinhoff {et~al.}(2016)Steinhoff, Hinderer, Buonanno, \&
  Taracchini}]{PhysRevD.94.104028}
Steinhoff, J., Hinderer, T., Buonanno, A., \& Taracchini, A. 2016, Phys. Rev.
  D, 94, 104028, \dodoi{10.1103/PhysRevD.94.104028}

\bibitem[{{Tan} {et~al.}(2020){Tan}, {Noronha-Hostler}, \& {Yunes}}]{Tan2020}
{Tan}, H., {Noronha-Hostler}, J., \& {Yunes}, N. 2020, arXiv e-prints,
  arXiv:2006.16296.
\newblock \doarXiv{2006.16296}

\bibitem[{{The LIGO Scientific Collaboration} \& {The Virgo
  Collaboration}(2020{\natexlab{a}})}]{userguide}
{The LIGO Scientific Collaboration}, \& {The Virgo Collaboration}.
  2020{\natexlab{a}}, LIGO/Virgo Public Alerts User Guide,
  https://emfollow.docs.ligo.org/userguide/

\bibitem[{{The LIGO Scientific Collaboration} \& {The Virgo
  Collaboration}(2020{\natexlab{b}})}]{gracedb}
---. 2020{\natexlab{b}}, Gravitational Wave Candidate Event DataBase,
  https://gracedb.ligo.org/

\bibitem[{{The LIGO Scientific Collaboration} \& {The Virgo
  Collaboration}(2020{\natexlab{c}})}]{GW19425samples}
---. 2020{\natexlab{c}}, Parameter estimation sample release for GW190425,
  https://dcc.ligo.org/LIGO-P2000026/public

\bibitem[{{The LIGO Scientific Collaboration} \& {The Virgo
  Collaboration}(20202)}]{GW190814samples}
---. 20202, GW190814 parameter estimation samples,
  https://dcc.ligo.org/LIGO-P2000183/public

\bibitem[{{Tsokaros} {et~al.}(2020){Tsokaros}, {Ruiz}, {Shapiro}, {Sun}, \&
  {Ury{\r{A}}'}}]{TsokarosRuiz2020}
{Tsokaros}, A., {Ruiz}, M., {Shapiro}, S.~L., {Sun}, L., \& {Ury{\r{A}}'}, K.
  2020, \prl, 124, 071101, \dodoi{10.1103/PhysRevLett.124.071101}

\bibitem[{Van~Oeveren \& Friedman(2017)}]{VanOeveren2017}
Van~Oeveren, E.~D., \& Friedman, J.~L. 2017, Phys. Rev. D, 95, 083014,
  \dodoi{10.1103/PhysRevD.95.083014}

\bibitem[{Vitale(2016)}]{Vitale2016}
Vitale, S. 2016, Phys. Rev. D, 94, 121501, \dodoi{10.1103/PhysRevD.94.121501}

\bibitem[{Weinberg(2016)}]{Weinberg_2016}
Weinberg, N.~N. 2016, The Astrophysical Journal, 819, 109,
  \dodoi{10.3847/0004-637x/819/2/109}

\bibitem[{{Wysocki} {et~al.}(2020){Wysocki}, {O'Shaughnessy}, {Wade}, \&
  {Lange}}]{Wysocki2020}
{Wysocki}, D., {O'Shaughnessy}, R., {Wade}, L., \& {Lange}, J. 2020, arXiv
  e-prints, arXiv:2001.01747.
\newblock \doarXiv{2001.01747}

\bibitem[{{Yang} {et~al.}(2018){Yang}, {East}, \& {Lehner}}]{YangEast2018}
{Yang}, H., {East}, W.~E., \& {Lehner}, L. 2018, \apj, 856, 110,
  \dodoi{10.3847/1538-4357/aab2b0}

\end{thebibliography}


\appendix

\section{Monte-Carlo Integrals}
\label{sec:monte carlo}

Some of our Monte-Carlo integral expressions are non-trivial, and as such we report them below.
We begin with the assumption that we have weighted sets of samples for \Mmax~and the mass of an individual object.
For example, given $x_i \sim p(x)$, we expect
\begin{equation}
    \frac{1}{N}\sum\limits_i^N \mathcal{F}(x_i) \approx \int dx\, p(x) \mathcal{F}(x)
\end{equation}
with $\mathcal{F}(x)$ an arbitrary function of $x$.
The basic quantity of interest in our calculation is
\begin{equation}
    P(m\leq M) = \frac{f}{g}
\end{equation}
where
\begin{gather}
    f = \frac{1}{N_M}\sum\limits_i^{N_M} p(\data_\EOS|M_i) \frac{1}{N_m}\sum\limits_k^{N_m} \frac{p(m_k|\Hyp)}{p(m_k|\Hyp_o)} \mathrm{H}(m_k \leq M_i) \label{eq:f} \\
    g = \frac{1}{N_M}\sum\limits_i^{N_M} p(\data_\EOS|M_i) \frac{1}{N_m}\sum\limits_k^{N_m} \frac{p(m_k|\Hyp)}{p(m_k|\Hyp_o)} \label{eq:g}
\end{gather}
assuming $M_i \sim p(M)$ and $m_k \sim p(m|\data, \Hyp_o)$ so that the integrals are done with respect to the measures $p(M|\data_\EOS) \propto p(M)p(\data_\EOS|M)$ and $p(m|\data,\Hyp) \propto p(m|\data, \Hyp_o) p(m|\Hyp)/p(m|\Hyp_o)$, respectively.
This approximates the types of Monte-Carlo integrals over the \EOS~realizations drawn from a prior done in \citet{Landry2020} as well as the process of reweighing single-event posteriors generated with one prior $\Hyp_o$ to match a different population prior $\Hyp$.
We investigate the behavior of these estimators and their correlated uncertainty from the finite number of Monte-Carlo samples.

Let us begin with $g$.
Its first and second moments under different realizations of sample sets are given by
\begin{align}
    \mathcal{E}[g] & = \int \prod\limits_i dM_i\, p(M_i) \prod\limits_k dm_k\, p(m_k|\data, \Hyp_o) \frac{1}{N_M}\sum\limits_j^{N_M} p(\data_\EOS|M_j) \frac{1}{N_m}\sum\limits_l^{N_m} \frac{p(m_l|\data, \Hyp)}{p(m_l|\data, \Hyp_o)} \nonumber \\
                   & = \left( \int dM\, p(M) p(\data_\EOS|M) \right) \left( \int dm\, p(m|\data, \Hyp_o) \frac{p(m|\Hyp)}{p(m|\Hyp_o)} \right) \nonumber \\
                   & = p(\data_\EOS) \left( \frac{p(\data|\Hyp)}{p(\data|\Hyp_o)} \right)
\end{align}
\begin{align}
    \mathcal{E}[g^2] & = \int \prod\limits_i dM_i\, p(M_i) \prod\limits_k dm_k\, p(m_k|\data, \Hyp_o) \left[ \frac{1}{N_M}\sum\limits_j^{N_M} p(\data_\EOS|M_j) \frac{1}{N_m}\sum\limits_l^{N_m} \frac{p(m_l|\Hyp)}{p(m_l|\Hyp_o)} \right]^2 \nonumber \\
                     & = \frac{1}{N_M N_m} \left( \int dM\, p(M) p(\data_\EOS|M)^2 \right) \left( \int dm\, p(m|\data, \Hyp_o) \left(\frac{p(m|\Hyp)}{p(m|\Hyp_o)}\right)^2 \right) \nonumber \\
                     & \quad\quad + \frac{N_m-1}{N_M N_m} \left( \int dM\, p(M) p(\data_\EOS|M)^2 \right) \left( \int dm\, p(m|\data, \Hyp_o) \frac{p(m|\Hyp)}{p(m|\Hyp_o)} \right)^2 \nonumber \\
                     & \quad\quad + \frac{N_M-1}{N_M N_m} \left( \int dM\, p(M) p(\data_\EOS|M) \right)^2 \left( \int dm\, p(m|\data, \Hyp_o) \left(\frac{p(m|\Hyp)}{p(m|\Hyp_o)}\right)^2 \right) \nonumber \\ 
                     & \quad\quad + \frac{(N_M - 1)(N_m - 1)}{N_M N_m} \left( \int dM\, p(M) p(\data_\EOS|M) \right)^2 \left(\int dm\, p(m|\data, \Hyp_o) \frac{p(m|\Hyp)}{p(m|\Hyp_o)} \right)^2
\end{align}
Similarly, we obtain
\begin{align}
    \mathcal{E}[f] & = \int dM dm\, p(M)p(\data_\EOS|M) p(m|\data, \Hyp_o) \frac{p(m|\Hyp)}{p(m|\Hyp_o)} \mathrm{H}(m \leq M)
\end{align}
\begin{align}
    \mathcal{E}[f^2] & = \frac{1}{N_M N_m} \int dM dm\, p(M)p(\data_\EOS|M)^2 p(m|\data, \Hyp_o) \left(\frac{p(m|\Hyp)}{p(m|\Hyp_o)}\right)^2 \mathrm{H}(m \leq M) \nonumber \\
                     & \quad\quad + \frac{N_m - 1}{N_M N_m} \int dM\, p(M) p(\data_\EOS|M)^2 \left( \int dm\, p(m|\data, \Hyp_o) \frac{p(m|\Hyp)}{p(m|\Hyp_o)} \mathrm{H}(m \leq M) \right)^2 \nonumber \\
                     & \quad\quad + \frac{N_M - 1}{N_M N_m} \int dm\, p(m|\data, \Hyp_o) \left(\frac{p(m|\Hyp)}{p(m|\Hyp_o)}\right)^2 \left( \int dM\, p(M) p(\data_\EOS|M) \mathrm{H}(m \leq M) \right)^2 \nonumber \\
                     & \quad\quad + \frac{(N_M - 1)(N_m - 1)}{N_M N_m} \left( \int dM dm\, p(M)p(\data_\EOS|M) p(m|\data, \Hyp_o) \frac{p(m|\Hyp)}{p(m|\Hyp_o)} \mathrm{H}(m \leq M) \right)^2
\end{align}
\begin{align}
    \mathcal{E}[fg] & = \frac{1}{N_M N_m} \int dM dm\, p(M) p(\data_\EOS|M)^2 p(m|\data, \Hyp_o) \left(\frac{p(m|\Hyp)}{p(m|\Hyp_o}\right)^2 \mathrm{H}(m \leq M) \nonumber \\ 
                    & \quad\quad + \frac{N_m - 1}{N_M N_m} \int dM\, p(M) p(\data_\EOS|M)^2 \left( \int dm\, p(m|\data, \Hyp_o) \frac{p(m|\Hyp)}{p(m|\Hyp_o)} \mathrm{H}(m \leq M) \right) \nonumber \\
                    & \quad\quad\quad\quad\quad\quad\quad\quad \times \left( \int dm\, p(m|\data, \Hyp_o) \frac{p(m|\Hyp)}{p(m|\Hyp_o)} \right) \nonumber \\
                    & \quad\quad + \frac{N_M - 1}{N_M N_m} \int dm\, p(m|\data, \Hyp_o) \left(\frac{p(m|\Hyp)}{p(m|\Hyp_o)}\right)^2 \left( \int dM\, p(M) p(\data_\EOS|M) \mathrm{H}(m \leq M)\right) \nonumber \\
                    & \quad\quad\quad\quad\quad\quad\quad\quad \times \left( \int dM\, p(M) p(\data_\EOS|M) \right) \nonumber \\
                    & \quad\quad + \frac{(N_M - 1)(N_m - 1)}{N_M N_m} \left(\int dM dm\, p(M) p(\data_\EOS|M) p(m|\data, \Hyp_o) \frac{p(m|\Hyp)}{p(m|\Hyp_o)} \mathrm{H}(m \leq M) \right) \nonumber \\
                    & \quad\quad\quad\quad\quad\quad\quad\quad \times \left( \int dM dm\, p(M) p(\data_\EOS) p(m|\data, \Hyp_o) \frac{p(m|\Hyp)}{p(m|\Hyp_o)} \right)
\end{align}
These expressions are exact, but as we may not be able to analytically integrate $p(M)p(\data_\EOS|M)$ or $p(m|\data, \Hyp)$, we approximate each with Monte-Carlo sums.
This implies the first moments are approximated by Eqs.~\ref{eq:f} and~\ref{eq:g}, while the second moments are approximately
\begin{align}
    \mathcal{E}[f^2] & \approx \frac{1}{N_M^2 N_m^2} \sum\limits_i^{N_M} \sum\limits_k^{N_m} p(\data_\EOS|M_i)^2 \left(\frac{p(m_k|\Hyp)}{p(m_k|\Hyp_o)}\right)^2 \mathrm{H}(m_k \leq M_i) \nonumber \\
                     & \quad\quad + \frac{N_m-1}{N_M^2 N_m^3} \sum\limits_i^{N_M} p(\data_\EOS|M_i)^2 \left( \sum\limits_k^{N_m} \frac{p(m_k|\Hyp)}{p(m_k|\Hyp_o)} \mathrm{H}(m_k \leq M_i) \right)^2 \nonumber \\
                     & \quad\quad + \frac{N_M-1}{N_M^3 N_m^2} \sum\limits_k^{N_m} \left(\frac{p(m_k|\Hyp)}{p(m_k|\Hyp_o)}\right)^2 \left( \sum\limits_i^{N_M} p(\data_\EOS|M_i) \mathrm{H}(m_k \leq M_i) \right)^2 \nonumber \\
                     & \quad\quad + \frac{(N_M-1)(N_m-1)}{N_M^3 N_m^3} \left( \sum\limits_i^{N_M} \sum\limits_k^{N_m} p(\data_\EOS|M_i) \frac{p(m_k|\Hyp)}{p(m_k|\Hyp_o)} \mathrm{H}(m_k \leq M_i) \right)^2
\end{align}
\begin{align}
    \mathcal{E}[g^2] & \approx \frac{1}{N_M^2 N_m^2} \left( \sum\limits_i^{N_M} p(\data_\EOS|M_i)^2 \right) \left( \sum\limits_k^{N_m} \left(\frac{p(m_k|\Hyp)}{p(m_k|\Hyp_o)}\right)^2 \right) \nonumber \\
                     & \quad\quad + \frac{N_m-1}{N_M^2 N_m^3} \left( \sum\limits_i^{N_M} p(\data_\EOS|M_i)^2 \right) \left( \sum\limits_k^{N_m} \frac{p(m_k|\Hyp)}{p(m_k|\Hyp_o)} \right)^2 \nonumber \\
                     & \quad\quad + \frac{N_M-1}{N_M^3 N_m^2} \left( \sum\limits_i^{N_M} p(\data_\EOS|M_i) \right)^2 \left( \sum\limits_k^{N_m} \left(\frac{p(m_k|\Hyp)}{p(m_k|\Hyp_o)}\right)^2 \right) \nonumber \\
                     & \quad\quad + \frac{(N_M-1)(N_m-1)}{N_M^3 N_m^3} \left( \sum\limits_i^{N_M} p(\data_\EOS|M_i) \right)^2 \left( \sum\limits_k^{N_m} \frac{p(m_k|\Hyp)}{p(m_k|\Hyp_o)} \right)^2
\end{align}
\begin{align}
    \mathcal{E}[fg] & \approx \frac{1}{N_M^2 N_m^2} \sum\limits_i^{N_M} \sum\limits_k^{N_m} p(\data_\EOS|M_i)^2 \left(\frac{p(m_k|\Hyp)}{p(m_k|\Hyp_o)}\right)^2 \mathcal{H}(m_k \leq M_i) \nonumber \\
                    & \quad\quad + \frac{N_m-1}{N_M^2 N_m^3} \sum\limits_i^{N_M} p(\data_\EOS|M_i)^2 \left( \sum\limits_k^{N_m} \frac{p(m_k|\Hyp)}{p(m_k|\Hyp_o)} \mathrm{H}(m_k \leq M_i) \right) \left( \sum\limits_l^{N_m} \frac{p(m_l|\Hyp)}{p(m_l|\Hyp_o)} \right) \nonumber \\
                    & \quad\quad + \frac{N_M-1}{N_M^3 N_m^2} \sum\limits_k^{N_m} \left(\frac{p(m_k|\Hyp)}{p(m_k|\Hyp_o)}\right)^2 \left( \sum\limits_i^{N_M} p(\data_\EOS|M_i) \mathrm{H}(m_k \leq M_i) \right) \left( \sum\limits_j^{N_M} p(\data_\EOS|M_j) \right) \nonumber \\
                    & \quad\quad + \frac{(N_M-1)(N_m-1)}{N_M^3 N_m^3} \left( \sum\limits_i^{N_M} \sum\limits_k^{N_m} p(\data_\EOS|M_i) \frac{p(m_k|\Hyp)}{p(m_k|\Hyp_o)} \mathrm{H}(m_k \leq M_i) \right) \nonumber \\
                    & \quad\quad\quad\quad\quad\quad\quad\quad \times \left( \sum\limits_i^{N_M} p(\data_\EOS|M_i) \right) \left( \sum\limits_k^{N_m} \frac{p(m_k|\Hyp)}{p(m_k|\Hyp_o)} \right)
\end{align}
Equipped with these uncertainty estimates, we approximate our statistics as follows
\begin{equation}
    P(m\leq M) = \frac{\mathcal{E}[f]}{\mathcal{E}[g]} \pm \sqrt{\left(\mathcal{E}[f^2]-\mathcal{E}[f]^2\right)\frac{1}{\mathcal{E}[g]^2} + \left(\mathcal{E}[g^2] - \mathcal{E}[g]^2\right)\frac{\mathcal{E}[f]^2}{\mathcal{E}[g]^4} - 2\left(\mathcal{E}[fg]-\mathcal{E}[f]\mathcal{E}[g]\right)\frac{\mathcal{E}[f]}{\mathcal{E}[g]^3}}
\end{equation}
\begin{align}
    \mathcal{O}^{m\leq M}_{m > M} & = \frac{P(m\leq M)}{1 - P(m\leq M)} \nonumber \\
                                    & = \frac{\mathcal{E}[f]}{\mathcal{E}[g] - \mathcal{E}[f]} \nonumber \\
                                    & \quad\quad \pm \sqrt{\left(\mathcal{E}[f^2]-\mathcal{E}[f]^2\right) \frac{\mathcal{E}[g]^2}{(\mathcal{E}[g]-\mathcal{E}[f])^4} + \left(\mathcal{E}[g^2] - \mathcal{E}[g]^2\right) \frac{\mathcal{E}[f]^2}{(\mathcal{E}[g]-\mathcal{E}[f])^4} - 2\left(\mathcal{E}[fg]-\mathcal{E}[f]\mathcal{E}[g]\right)\frac{\mathcal{E}[g]\mathcal{E}[f]}{(\mathcal{E}[g]-\mathcal{E}[f])^4}}
\end{align}

We note that similar error estimates are possible for $\mathcal{B}^{m\leq M}_{m\geq M}$ via Monte-Carlo approximates to
\begin{equation}
    \mathcal{B}^{m\leq M}_{m > M} = \left(\frac{f}{g-f}\right)\left(\frac{\mathcal{G}-\mathcal{F}}{\mathcal{F}}\right)
\end{equation}
where
\begin{gather}
    \mathcal{F} = \frac{1}{N_M}\sum\limits_i^{N_M} p(\data_\EOS|M_i) \frac{1}{N_p}\sum\limits_j^{N_p} \frac{p(m_j|\Hyp)}{p(m_j|\Hyp_o)} \mathrm{H}(m_j \leq M_i) \label{eq:f} \\
    \mathcal{G} = \frac{1}{N_M}\sum\limits_i^{N_M} p(\data_\EOS|M_i) \frac{1}{N_p}\sum\limits_j^{N_p} \frac{p(m_j|\Hyp)}{p(m_j|\Hyp_o)} \label{eq:g}
\end{gather}
where the $N_p$ mass samples are drawn from the single-event prior ($m_k\sim p(m|\Hyp_o)$) instead of the posterior.
This follows from separate estimates from the posterior odds and prior odds in Eqs.~\ref{eq:odds ratio} and~\ref{eq:decomposed odds}.
We use a similar decomposition into separate (correlated) Monte-Carlo estimates for the numerator and denominator of the posterior and prior odds separately as well as cross terms from the fact that we use the same set of \Mmax~samples within both estimates.
However, as we do not report $\mathcal{B}^{m\leq M}_{m > M}$ within our analysis, we leave the technical details as an exercise for the reader.

\end{document}